%% file: ICDE_CR/main.tex
\documentclass[conference]{IEEEtran}
\IEEEoverridecommandlockouts
\usepackage{cite}
\usepackage{amsmath,amssymb,amsfonts}
\usepackage{graphicx}
\usepackage{textcomp}
\usepackage{xcolor}
\usepackage{blindtext}
\usepackage[normalem]{ulem}
\usepackage{graphicx} 
\usepackage{url}
\usepackage{mathtools}
\usepackage{capt-of} 

\usepackage{xcolor} 
\usepackage{marginnote}

\usepackage[linesnumbered,ruled,vlined]{algorithm2e}
\usepackage[hidelinks]{hyperref} 
\usepackage{orcidlink}
\usepackage{amsmath}
\usepackage{booktabs}
\usepackage{colortbl}
\usepackage{enumerate}
\usepackage[mathscr]{eucal}
\usepackage{graphicx}
\usepackage{makecell}
\usepackage{mathtools}
\usepackage[font=rm]{caption}
\DeclareCaptionType{copyrightbox}
\usepackage{shortvrb}
\usepackage{tabularx}
\usepackage{threeparttable}
\usepackage{verbatim}
\usepackage{xspace}
\usepackage{tcolorbox}
\usepackage{caption}
\usepackage{subcaption}
\usepackage{listings}
\usepackage{fancyvrb}
\usepackage{enumitem}
\usepackage{multirow}
\usepackage{amsmath}
\usepackage{bbding}
\usepackage{siunitx}
\usepackage{bm}
\usepackage{algpseudocode}
\usepackage{textcomp}
\usepackage{xparse}
\setlist[itemize,1]{leftmargin=3mm,itemsep=0mm}
\setlist[enumerate,1]{leftmargin=5mm,itemsep=0mm}
\usepackage{balance}
\usepackage{makecell}
\usepackage{xspace}
\usepackage{ragged2e}
\newif\ifcameraready
\camerareadyfalse

\def\BibTeX{{\rm B\kern-.05em{\sc i\kern-.025em b}\kern-.08em
    T\kern-.1667em\lower.7ex\hbox{E}\kern-.125emX}}
\begin{document}

\input{ICDE_MTQA/macros}

\title{Decomposition-Driven Multi-Table Retrieval and Reasoning for Numerical Question Answering}

\author{
\IEEEauthorblockN{%
Feng Luo\textsuperscript{1}\textsuperscript{\textdagger},
Hai Lan\textsuperscript{2},
Hui Luo\textsuperscript{3},
Zhifeng Bao\textsuperscript{2},
Xiaoli Wang\textsuperscript{4},
J.\,Shane Culpepper\textsuperscript{2},
Shazia Sadiq\textsuperscript{2}%
\thanks{\textsuperscript{\textdagger}This work was done while visiting The University of Queensland.}
}
\IEEEauthorblockA{%
\textsuperscript{1}RMIT University \quad
\textsuperscript{2}The University of Queensland \quad
\textsuperscript{3}University of Wollongong \quad
\textsuperscript{4}Xiamen University\\
feng.luo@student.rmit.edu.au,\;
h.lan@uq.edu.au,\;
huil@uow.edu.au,\;
zhifeng.bao@uq.edu.au,\\
xlwang@xmu.edu.cn,\;
s.culpepper@uq.edu.au,\;
shazia@eecs.uq.edu.au%
}
}

\maketitle

\begin{abstract}
\input{ICDE_MTQA/0_abstract}
\end{abstract}


\input{ICDE_MTQA/1_intro_new}

\input{ICDE_MTQA/2_problem}
\input{ICDE_MTQA/3_method_qa}

\input{ICDE_MTQA/3_method_retriever}

\input{ICDE_MTQA/3_method_reasoner}

\input{ICDE_MTQA/4_0_dataset_preparation}
\input{ICDE_MTQA/4_experiment}

\input{ICDE_MTQA/5_conclusion}

\bibliographystyle{IEEEtran}
\bibliography{ICDE_CR/refer}

\clearpage
\appendices 
\input{ICDE_MTQA/6_appendix}

\end{document}

%% file: ICDE_MTQA/macros.tex
\newcommand{\bao}[1]{{\color{brown}{\bf{Bao says:}} \emph{#1}}}
\newcommand{\hui}[1]{{\color{magenta}{\bf{Hui says:}} \emph{#1}}}
\newcommand{\feng}[1]{{\color{blue}{\bf{Feng says:}} \emph{#1}}}
\newcommand{\shane}[1]{\textrm{\textcolor{orange}{Shane says: #1}}}
\newcommand{\wxl}[1]{{\color{pink}{\bf{WXL says:}} \emph{#1}}}
\newcommand{\revise}[1]{{\color{black}#1}}
\newcommand{\hlight}[1]{{\color{black}#1}}

\newcommand{\mn}[1]{\ensuremath{\mathnormal{#1}}}
\newcommand{\mnth}[1]{\ensuremath{\mathnormal{#1}\cdot}th}
\newcommand{\svar}[1]{\mbox{\scriptsize\emph{#1}}}
\newcommand{\tvar}[1]{\mbox{\tiny\emph{#1}}}
\newcommand{\avar}[1]{\mbox{#1}}
\newcommand{\asvar}[1]{\mbox{\scriptsize{#1}}}
\newcommand{\atvar}[1]{\mbox{\tiny{#1}}}
\def\D{\hphantom{1}}
\def\C{\hphantom{1,}}
\newcommand{\metricfont}[1]{{\small\sf{#1}}}
\newcommand{\metric}[1]{\metricfont{#1}}

\newcommand{\pp}{\textcolor{cyan}}
\newcommand{\update}[1]{{\color{black}#1}}
\newcommand{\newpar}{\noindent}
\newcommand{\ourmethod}{\textsf{DMRAL}}
\newcommand{\methodqa}{Table-Aligned Question Decomposer}
\newcommand{\methodrt}{Coverage-Aware Retriever}
\newcommand{\methodrs}{Sub-question Guided Reasoner}

\newcommand{\graphstructure}{\textit{Table Relationship Graph}}

\newcommand{\settingdef}{large-scale table collections}

\newtheorem{thm}{Theorem}
\newtheorem{lem}{Lemma}
\newtheorem{defn}{Definition}
\newtheorem{prob}{Problem}
\newtheorem{ex}{Example}

\setlength{\marginparwidth}{1cm}  
\setlength{\marginparsep}{3pt}     

\reversemarginpar
\renewcommand{\marginnote}[2][]{}

\newcommand{\revnote}[1]{%
  \marginpar{\raggedright\textcolor{blue}{#1}}%
}

%% file: ICDE_MTQA/0_abstract.tex
In this paper, we study the problem of numerical multi-table question answering (MTQA) over \settingdef{} (e.g., online data repositories).
This task is essential in many analytical applications. 
Existing MTQA solutions, such as text-to-SQL or open-domain MTQA methods, are designed for databases and struggle when applied to \settingdef{}.
The key limitations include:
(1) Limited support for complex table relationships; (2) Ineffective retrieval of relevant tables at scale; (3) Inaccurate answer generation.
To overcome these limitations, we propose \ourmethod{}, a \underline{D}ecomposition-driven \underline{M}ulti-table \underline{R}etrieval and \underline{A}nswering framework for MTQA over \underline{L}arge-scale table collections, which consists of:
(1) constructing a table relationship graph to capture complex relationships among tables;
(2) \methodqa{} and \methodrt{}, which jointly enable the effective identification of relevant tables from large-scale corpora by enhancing the question decomposition quality and maximizing the question coverage of retrieved tables; 
(3) \methodrs{}, which produces correct answers by progressively generating and refining the reasoning program based on sub-questions.
Experiments on two MTQA datasets demonstrate that \ourmethod{} significantly outperforms existing state-of-the-art MTQA methods, with an average improvement of 24\% in table retrieval and 55\% in answer accuracy.

%% file: ICDE_MTQA/1_intro_new.tex
\section{Introduction} \label{intro}

Multi-table question answering (MTQA) is a well-known task that requires identifying and integrating information from multiple tables to answer a single question \cite{wummqa}.
\update{Among these questions, those requiring \textit{numerical answers} exhibit substantially greater complexity than those seeking \textit{textual answers}. Empirical evidence indicates a pronounced performance gap: approximately $55\%$ accuracy for numerical answers vs. $88\%$ for textual answers, even with gold tables \cite{qiu2024tqa,pal2023multitabqa}, which underscores the imperative for MTQA systems to accurately support numerical questions~\cite{nararatwong2024dbqr}.}
At the same time, the growing prevalence of \revise{large-scale table collections}—such as tables found on the web, in public data lakes, or available through data markets—offers rich opportunities for MTQA, where the valuable data for answering the numerical questions is often dispersed across those soiled tables.
While promising, \hlight{working with such tables presents unique challenges due to their \emph{large scale} (e.g., tens of thousands of tables), potentially \emph{incomplete metadata} (e.g., missing column headers) \cite{solo2023sigmod,khatiwada2023santos}, and \emph{complex inter-table relationships}, namely unionability (tables that can be unioned on similar column headers) and joinability (tables that can be joined based on matching columns) \cite{fan2022semantics, dong2023deepjoin}}. 
In this paper, we study numerical MTQA over \settingdef{}, a challenging but increasingly realistic scenario driven by real-world data acquisition and usage trends.

\begin{figure*}[t]
    \centering
    \includegraphics[width=0.9\linewidth]{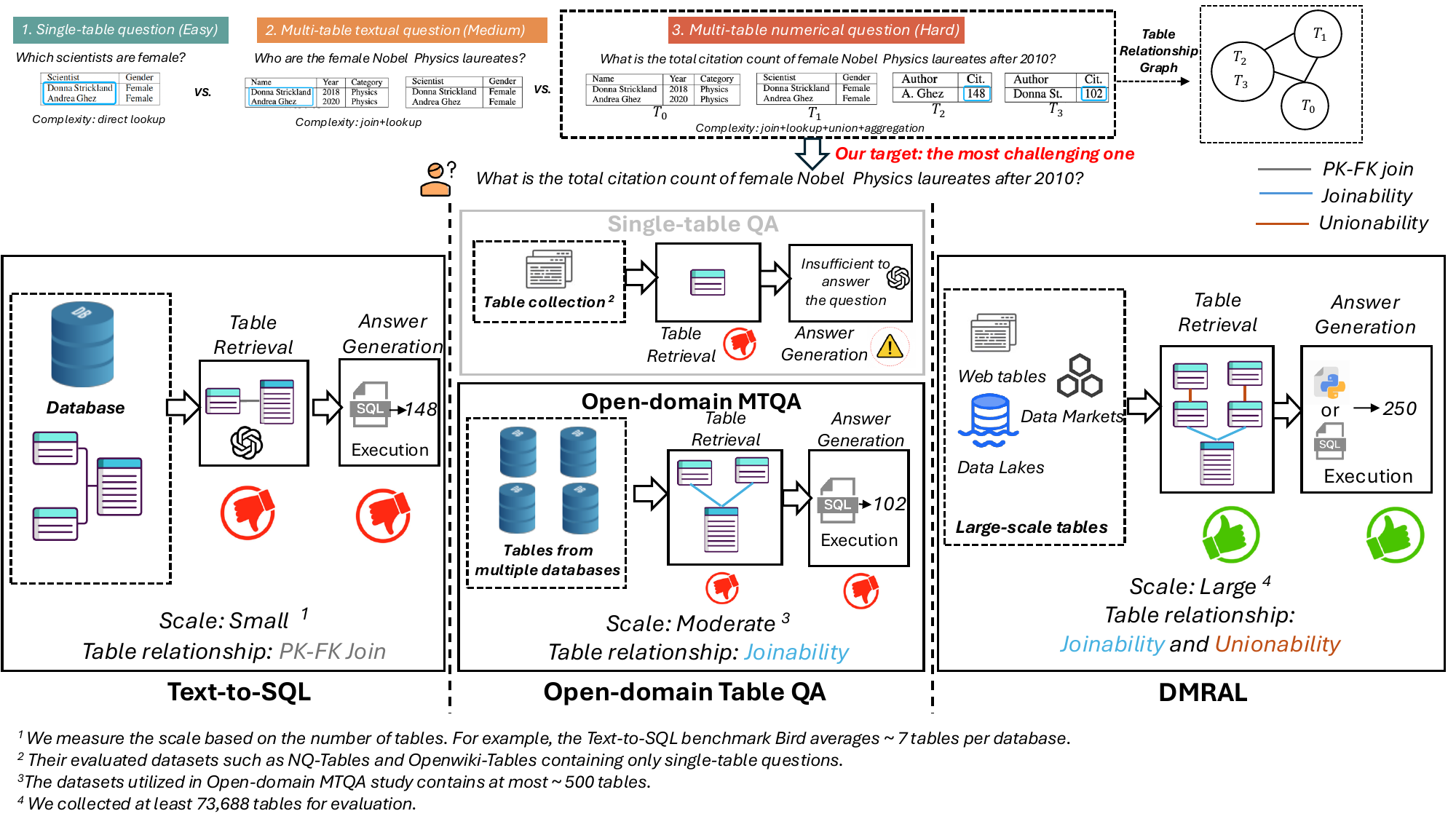}
    \vspace{-3mm}
    \caption{A comparison of problem settings and system capabilities among existing MTQA approaches and our proposed DMRAL.}
    \label{sec_intro_comparison_figure}
    \vspace{-3mm}
\end{figure*}

\update{Current approaches to MTQA broadly fall into two categories: \emph{Text-to-SQL} and \emph{Open-domain Table QA}. 
As shown in Figure \ref{sec_intro_comparison_figure}, \hlight{Text-to-SQL methods~\cite{ li2025alpha, mohammadjafari2024natural} were originally developed for a single relational database, typically involving fewer than ten tables and relying on a well-defined database schema (i.e., complete table metadata and PK-FK constraints \cite{chen2024beaver}).
Such schema information and limited scale are not available in our problem, which makes Text-to-SQL methods not directly applicable, as we will describe in \S \ref{sec_related_work} shortly. 
}
Open-domain Table QA is further divided into (i) single-table QA~\cite{herzig2021open,pan2022end}, which assumes the answer requires information solely from a single table and is therefore incompatible with our setting, and (ii) open-domain MTQA~\cite{chen2024table,wummqa}, which combines evidence from multiple tables.
While open-domain MTQA is closer to our problem, it is designed for modest corpora (i.e., hundreds of tables) that are aggregated from multiple relational databases. 
These approaches typically begin by decomposing a question into a set of sub-questions using the internal knowledge of LLMs, then retrieving relevant tables based on a relevance score between sub-questions and tables using table joinability, and finally invoking an LLM to generate SQL over the retrieved tables.
}

However, applying existing open-domain MTQA methods to large-scale table collections presents three major limitations.\hlight{
\noindent\textbf{(L1)}~\textit{Limited support for complex table relationships}: Open-domain MTQA
overlooks complex table relationships such as unionability \cite{zhu2024autotqa}, which limits their robustness in handling such table corpora.
\noindent\textbf{(L2)}~\textit{ Ineffective retrieval of relevant tables at scale}: 
Open-domain MTQA depends on LLMs to decompose sub-questions for table retrieval, while the errors from low-quality decompositions can propagate to retrieval, leading to suboptimal retrieval effectiveness.
\noindent\textbf{(L3)}~\textit{Inaccurate answer generation}: While generating programs such as SQL or Python and running them to derive answers is promising for answering numerical questions, existing approaches rarely produce fully correct programs (e.g., including incorrect joins), resulting in low answer accuracy \cite{DBLP:journals/corr/abs-2501-09310}.
}

\smallskip
\noindent \textbf{Our Contributions.}  
To resolve these limitations, we propose \ourmethod{}, a \underline{D}ecomposition-driven \underline{M}ulti-table \underline{R}etrieval and \underline{A}nswering framework for MTQA over \underline{L}arge-scale table collections.
Our key contributions are as follows:
\begin{itemize}[leftmargin=*,noitemsep,topsep=0pt,partopsep=0pt,parsep=0pt]
    \item We develop a \textit{Preprocessing} pipeline (\S\ref{sec_problem_solution_overview}) that builds a \graphstructure{} to capture complex table relationships (i.e., addressing \textbf{L1}).
    \item \hlight{We propose an effective decomposition-driven multi-table retrieval strategy to address \textbf{L2}, consisting of: (1) \textit{\methodqa{}} (\S\ref{sec_method_qd}), which enhances decomposition quality by identifying distinct information needs and aligning them with the underlying table structures; (2) \textit{\methodrt{}} (\S\ref{sec_method_rt}), which effectively retrieves relevant tables at scale and mitigates cascading retrieval errors by maximizing their coverage of the question through coverage scoring and verification mechanisms.}
    \item \hlight{We propose a \textit{\methodrs{}} (\S\ref{sec_method_rs}) to improve answer accuracy by guiding LLMs to progressively generate and refine the program based on the decomposed sub-questions (i.e., addressing \textbf{L3}).} 
    \item To comprehensively evaluate this task, we prepare two large-scale datasets, SpiderWild and BirdWild, consisting of 73,688 and 109,949 tables curated from real-world table sources~\cite{yu2018spider,li2023can,deng2024lakebench} (\S \ref{sec:data_prep}). Extensive experiments demonstrate the effectiveness of \ourmethod{}, achieving an average improvement of 24\% in identifying relevant tables and 55\% in producing accurate answers (\S\ref{sec_exp_total}).
\end{itemize}
\vspace{-0.6em}
In summary, \ourmethod{} is a robust, scalable, and traceable framework designed for numerical MTQA over \settingdef{}. 
\ourmethod{} enables fine-grained tracing and verification (e.g., whether the retrieved tables are appropriate, whether the reasoning program is correct, and whether the sub-questions are well-decomposed). This traceability is crucial, as it makes answer derivations transparent and provides actionable insights for diagnosing and improving each component.

%% file: ICDE_MTQA/2_problem.tex
\section{Problem, Literature, and Solution Overview}
\subsection{Problem Definition}
A table $T$ consists of a set of columns $T_c = \{c_1, \dots, c_m\}$ and rows $T_r$, along with associated metadata. 
The metadata includes a table title $T_p$ and column headers $T_h = \{h_1, \dots, h_m\}$, which may be partially missing.
We define the table collection as $\mathcal{T} = \{T^1, T^2, \dots, T^{|\mathcal{T}|}\}$. 
These tables are often related through two representative inter-table relationships:
\begin{itemize}
    \item \textbf{Joinability:} Two tables $T^i$ and $T^j$ are considered joinable if there exists at least one column from $T_c^i$ and one from $T_c^j$ such that the two columns share overlapping or semantically similar values.
    
    \item \textbf{Unionability:} Two tables $T^i$ and $T^j$ are considered unionable if their column headers $T_h^i$ and $T_h^j$ are sufficiently similar to allow a one-to-one alignment across two tables.
\end{itemize}

\begin{defn} [Numerical Multi-Table Question Answering]
    Given a table collection $\mathcal{T}$ and a numerical question $q$ that requires a numeric answer, it aims to compute the correct answer $ans$ by learning the function $f: (\mathcal{T}, q) \rightarrow ans$, where the answer $ans$ is derived by identifying a subset of relevant tables $\mathcal{T}_q \subseteq \mathcal{T}$ and performing reasoning over $\mathcal{T}_q$.
\end{defn}

\subsection{Related Work} \label{sec_related_work}

\noindent \textbf{Text-to-SQL.}
Text-to-SQL is a long-standing task that aims to convert natural language (NL) questions into executable SQL queries over relational databases.
Existing Text-to-SQL approaches fall into three paradigms: rule-based, neural, and LLM-based.
Rule-based methods use hand-crafted grammars and semantic parsers~\cite{li2014sigmod,fu2023pvldb,yu2018syntaxsqlnet}, but they require heavy manual engineering and generalize poorly to new domains~\cite{liu2025survey}.
Neural methods leverage deep models ~\cite{xiao2016acl, yin2020acl,bogin2019acl}
to improve schema understanding, yet remain limited by model capacity and the availability of high-quality training data.
LLM-based methods generate SQL via prompt engineering~\cite{pourreza2023din,xie2025opensearchsqlenhancingtexttosqldynamic}, supervised fine-tuning~\cite{li2024pvldb}, and tool-augmented/agentic reasoning~\cite{chen2025pacmmod,actsql,li2025alpha,talaei2024chess}.
For example, CHESS~\cite{talaei2024chess} proposes a multi-agent pipeline that stages schema selection, SQL generation, and validation. OpenSearch-SQL~\cite{xie2025opensearchsqlenhancingtexttosqldynamic} develops a multi-agent collaboration approach with four stages: Preprocessing, Extraction, Generation, and Refine, achieving state-of-the-art results~\cite{liu2025survey}.

Despite their success, these methods are not directly applicable to MTQA over \settingdef{}, due to two key limitations.
First, they assume explicit table relationships via PK-FK joins~\cite{furst2024evaluating} and require complete metadata for schema linking~\cite{wang2020rat}, both of which are unavailable in real-world table collections.
Second, the SOTA LLM-based methods are limited by the context length of LLMs, and do not scale for large-scale table corpora~\cite{wu2024tablebench}.
\revise{
Given these limitations, we do not consider Text-to-SQL methods as primary baselines, while we provide two complementary comparisons: (i) an adapted evaluation where we run Text-to-SQL over the same table collections via database materialization, and (ii) a secondary evaluation under their original settings (\S \ref{sec_exp_text2sql_01}).
}

\smallskip
\noindent \textbf{Open-domain MTQA.}
Open-domain MTQA operates over table collections by first retrieving the relevant tables using the decomposed sub-questions, and then generating SQL queries directly using LLMs \cite{zhang2023nameguess}.
To identify the relevant tables, MMQA \cite{wummqa} formulates the problem as a multi-hop retrieval problem.
In contrast,  JAR~\cite{chen2024table} formulates the problem as an optimization problem by jointly considering sub-question–table relevance and overall table relevance.
However, both approaches are designed for database tables, which limits their ability to support \update{identifying unionable tables}.
Moreover, MMQA is vulnerable to cascading errors introduced by early retrieval mistakes~\cite{lee2022generative}, reducing its \textit{effectiveness}.
Conversely, JAR suffers from a high computational overhead due to the optimization algorithm used, limiting \textit{efficiency}.

\smallskip
\update{\noindent \textbf{Other loosely related work.}
Open-domain Single-Table QA~\cite{herzig2021open,pan2022end,yu2025tablerag} aims to answer NL questions using only the most relevant table retrieved from a collection of diverse tables, relying on dedicated single-table retrieval solutions such as \cite{balaka2025pneuma} without considering inter-table relationships.
This fundamental assumption makes them incompatible for integrating information from multiple tables to answer a question.

\begin{figure}[t]
\centering
\includegraphics[width=0.46\textwidth]{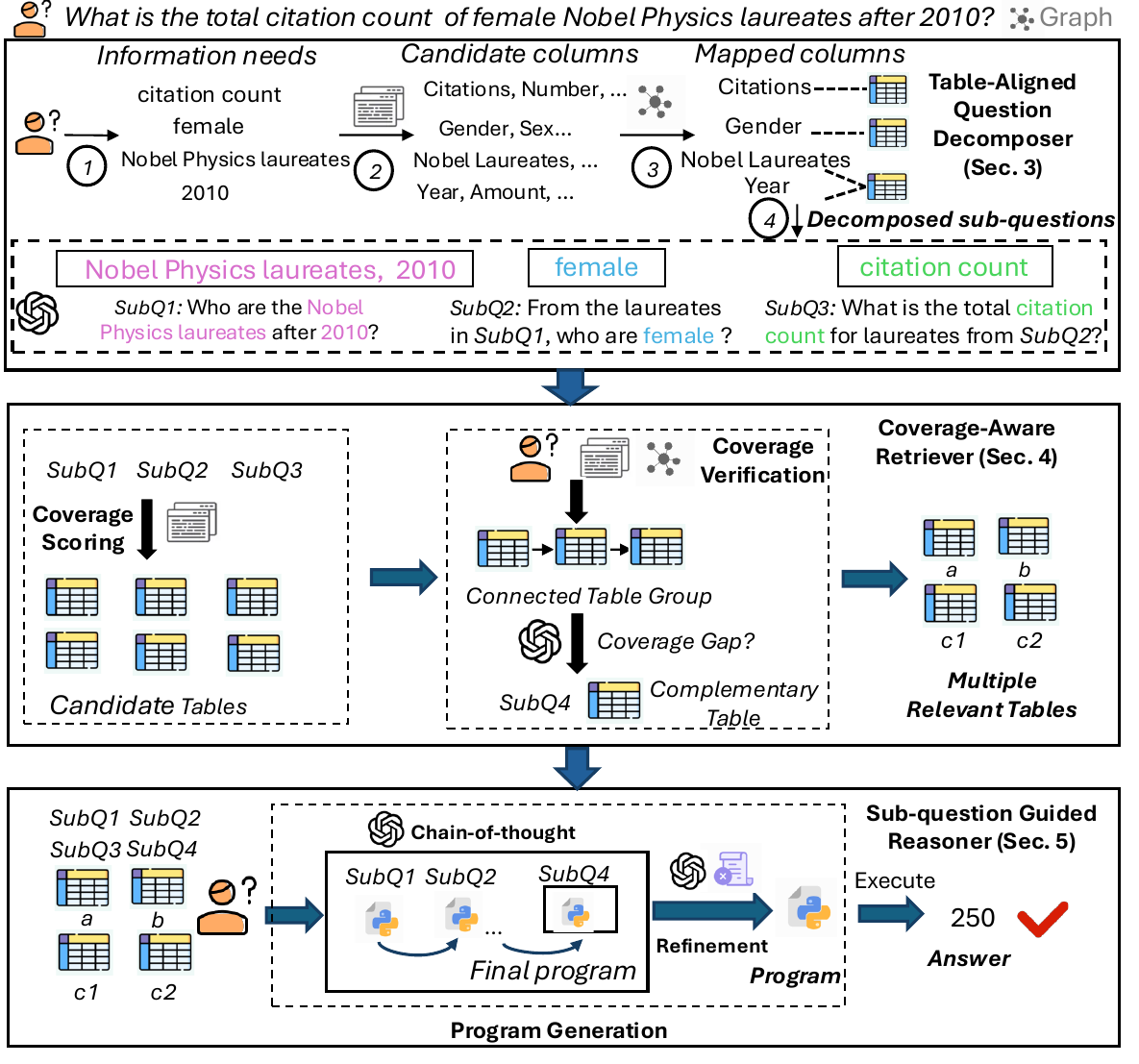}
\vspace{-3mm}
\caption{MTQA processing with \ourmethod{}, in which intermediate results at each step can be traced and refined.}
\label{fig:solution_overview}
\vspace{-5mm}
\end{figure}
}
\subsection{Solution Overview} \label{sec_problem_solution_overview}
\update{\ourmethod{} begins with a \textit{Preprocessing} pipeline that constructs a \graphstructure{} $\mathcal{G} = (\mathcal{V}, \mathcal{E})$ to efficiently capture complex table relationships.  
In this graph, each node $v \in \mathcal{V}$ represents a cluster of unionable tables, identified using the unionability from \cite{sarma2012finding}.
Edges $e \in \mathcal{E}$ connect these clusters if any pair of tables in the clusters is joinable, using joinability defined in \cite{chen2024table}. } 
Built upon this preprocessed graph\revise{(see the concrete example in Figure~\ref{sec_intro_comparison_figure})}, \ourmethod{} answers questions using three core modules, as illustrated in Figure~\ref{fig:solution_overview}:

\noindent \textbf{\methodqa{}} (\S\ref{sec_method_qd}) decomposes the input question into multiple sub-questions.
To improve decomposition quality, it proposes a four-step decomposition approach: (1) extract the information needs from the question, (2) align each information need to the columns, (3) select the most promising mapping between information needs and columns, and (4) group contextually coherent information needs to generate subquestions using LLM-based decomposition.

\noindent \textbf{\methodrt{}} (\S\ref{sec_method_rt}) retrieves multiple relevant tables based on the decomposed subquestions.
To support effective retrieval over large scale of tables, it first introduces a learning-based coverage scoring function to retrieve candidate tables for each sub-question.
Then it proposes a coverage verification function to first construct the connected table group to collectively answer the question, and then identify complementary tables to fill any potential coverage gap.

\noindent \textbf{\methodrs{}} (\S\ref{sec_method_rs}) generates executable programs (e.g., SQL or Python) over the retrieved tables to derive the final answer.
To improve answer reliability, it first uses chain-of-thought prompting to incrementally generate intermediate programs using a sequence of decomposed sub-questions.
It then applies an execution-guided refinement mechanism to verify and iteratively revise the generated program.

%% file: ICDE_MTQA/3_method_qa.tex
\section{Table-Aligned Question Decomposer} \label{sec_method_qd}

\subsection{Principles of Effective Decomposition}
\textit{Correct decomposition of complex user questions} is essential for MTQA, as it enables precise retrieval \cite{chen2024table}.
Given a question $q$, we define the information needs contained as a set $\mathcal{I}(q)$, where each element $I_q \in \mathcal{I}(q)$ represents a specific concept, entity, or condition included in $q$ that is required to derive the correct answer.

We argue that high-quality question decomposition should satisfy these three guiding principles: 
(1) \textbf{Completeness}: The generated sub-questions must collectively cover all of the information needs contained in $q$.
Missing any aspect of a question may result in an incomplete answer that fails to fully meet the user’s intent; 
(2) \textbf{Non-redundancy}: There should be no overlap between the information needs in each sub-question. 
Redundant information requires unnecessary computation, increases retrieval latency, and can create misleading input for the reasoning; 
(3) \textbf{Table-specificity}: Each sub-question targets a single table or a group of unionable tables. 
This reduces the complexity of multi-table joins and ensures that each sub-question is answerable within a coherent context.

While LLMs have shown promise for decomposition~\cite{zhang2024tree, peng2024chain, wummqa} and improved decomposition accuracy compared to previous work~\cite{min2019multi,huang2023question}, our empirical analysis (\S \ref{sec_exp_ablation}) reveals that simply using LLMs for decomposition often fails to meet the above principles, leading to incorrect answers.
\subsection{From Question to Table-Specific Sub-questions}
To address these limitations and satisfy the principles outlined above, we propose a four-step decomposition approach that consistently aligns sub-questions with the structural information of the table collection.

\subsubsection{Identifying Information Needs}\label{sec_ident_info_needs}
\hlight{ To extract the most salient information needs from each question, 
we first parse a question into a constituency tree using the Stanford Stanza toolkit \cite{stanza}.
Then,  we extract all the noun phrases, which represent the core concepts and entities required, and include adjective and verb phrases that may represent the conditions, based on the syntactic labels~\cite{lan2024ps}.
This produces a comprehensive set of ``information needs'' to ensure \textbf{Completeness}.}

\subsubsection{Hybrid Column Matching} \label{sec_method_htm}



To ensure alignment between information needs and columns, we construct a text snippet to represent each column $c_j$ in table $T$ by concatenating the table title $T_p$, column header $h_j$, and distinct cell values from $c_j$, separated by spaces. 
Since information needs can differ from table contents both lexically and semantically, \hlight{we use M3-Embedding (M3)~\cite{chen2024m3}, a unified embedding model designed to encode both lexical and semantic features of text. }
We then encode all column snippets in the table collection $\mathcal{T}$ using M3 \hlight{and index them with FAISS \cite{douze2024faiss} to support efficient similarity search}. 
For each information need $I_q$, we encode an M3 embedding and retrieve the top-$30$ column snippets based on their similarity scores.
A retrieval depth of $30$ was chosen empirically to balance relevance against noise (irrelevant columns).


\subsubsection{Context-Aware Column Disambiguation}\label{sec_method_ctd}
For each information need $I_q$, we identify the candidate columns $\mathcal{C}_{I_q}$, along with their similarity scores $\text{sim}(I_q, c_{I_q})$ for each $c_{I_q} \in \mathcal{C}_{I_q}$.
To guide sub-question generation, we select the most promising mapping between the information needs and columns to ensure contextual coherence across information needs.

Since the information needs for a question $q$ may be interrelated, they should refer to columns from a single table or from a set of joinable tables. 
To achieve this, we use the  \revise{Table Relationship Graph} $\mathcal{G}$ that contains any required contextual knowledge, and define a ``contextual relevance score'' that jointly scores all candidate columns for each information need.

\noindent \textbf{Contextual Relevance Score}. Given a specific mapping $M$ that assigns a single column from $\mathcal{C}_{I_{q}}$ to $I_q$ for each $I_q$ $\in \mathcal{I}_q$, the {\em context relevance score} $R(M,\mathcal{I}_q)$ is defined as $\sum_{(I_q, c_{I_q})\in M}\text{sim}(I_q,c_{I_q})$ if $\mathbb{I} \left( \{ T(c_{I_q}) \mid I_q \in \mathcal{I}(q) \}, \mathcal{G} \right)= 1$. 
Otherwise, $R(M,\mathcal{I}_q) = 0$. 
Here, $T(c_{I_q})$ denotes the table containing column $c_{I_q}$, and $\mathbb{I}(\cdot, \mathcal{G})$ is an indicator function returning 1 if all selected tables belong to nodes forming a connected component in $\mathcal{G}$. 

Our objective is to identify a mapping $M$ that achieves the highest contextual relevance score for all possible mappings. 
However, the number of candidate mappings grows exponentially with the number of information needs, which can be large. 
So, exhaustive search is not computationally efficient.
Therefore, we apply a greedy strategy on $\mathcal{G}$. 
The strategy consists of the following steps.

\noindent\textit{\underline{Step 1: Ranking Information Needs.}}
We rank all information needs in a descending order based on the maximal similarity  between each information need and its candidate columns. 

\noindent\textit{\underline{Step 2: Initializing the Mapping.}}
We initialize the mapping $M$ by selecting the candidate column with the highest similarity score for the top-ranked information need.

\noindent\textit{\underline{Step 3: Expanding the Mapping Progressively.}}
We iterate through the remaining sorted information needs one by one. 
For each information need, we select the candidate column that (1) belongs to a table that is from the same connected component in $\mathcal{G}$ with the tables already included in the current mapping $M$, and (2) 
with the highest similarity score among these connected candidates.
If no candidate satisfies these requirements, the current expansion is considered unsuccessful.
In such cases, we backtrack to Step 2 to retry the process by selecting the next-best candidate column based on the similarity scores.





\subsubsection{Question Decomposition}\label{sec_ques_decom}
Once the optimal column selection is obtained, we group the information needs based on the tables containing the columns selected.
These table-aligned groups are then sent to an LLM to generate a single sub-question for each group. 
This step \textbf{reduces redundancy} by ensuring that each sub-question corresponds to a disjoint set of targeted needs, and leads to more consistent \textbf{Table-specificity} by limiting the scope of each sub-question.

%% file: ICDE_MTQA/3_method_retriever.tex
\section{Coverage-Aware Retriever} \label{sec_method_rt}



\textit{Retrieving complete and relevant sets of tables} is a key requirement for accurate multi-table question answering.
However, existing approaches either incur high computational costs~\cite{chen2024table} or include overly rigid multi-hop pipelines that increase error rate~\cite{wummqa}. 
To address this, we introduce Coverage-Aware Retriever to support more efficient multi-hop retrieval, which relies on two key innovations:
(1) A coverage scoring function that prioritizes question semantic coverage to improve the effectiveness of early-stage candidate selection (\S \ref{solution_coverage_scoring}); (2) A coverage verification function that detects and corrects any missing information using {\em residual sub-questions} (\S \ref{sec:coverage_verification}).



\subsection{Maximizing Coverage via Learning-based Scoring} \label{solution_coverage_scoring}

Our approach uses a two-stage pipeline. 
First, we perform candidate retrieval using FAISS and M3 table embeddings. 
Then, we apply a more fine-grained reranking model to increase the precision.

\noindent\textbf{Coarse Retrieval}. We represent each table cluster in $\mathcal{G}$ as a document by concatenating the table metadata and encoding it using M3. 
The embeddings produced are then indexed using FAISS. 
\revise{ We avoid embedding full content, given that flattening all content can exceed the encoder context limits for large tables (e.g., with $>1$K rows), and introduce spurious matches due to frequent entities (e.g., years).
}
For each sub-question ${sq}_i$, we retrieve all candidate clusters by querying the index.

\hlight{
\noindent\textbf{Learned Scoring.} Coarse retrieval often introduces irrelevant documents due to superficial semantic similarity, which can propagate errors in later stages. 
To increase precision during candidate selection, we train a scoring function $f_\theta(q, T_q)$ that estimates the \emph{semantic coverage} of a candidate table $T_q$ w.r.t a question $q$. 
We use ColBERTv2~\cite{santhanam2022colbertv2} to create $f_\theta$ due to its effectiveness in capturing contextual interactions.}
\revise{ $T_q$ is represented using the same document with compact table metadata, for scalability under ColBERTv2's context limits.}

To construct the training data, we include single-table QA datasets~\cite{yu2018spider,li2023can}. 
Each training instance consists of a question $q$, a positive table $T^+$ (i.e., the ground-truth table containing the correct answer), and a hard negative table $T^-$ created by removing the answer-bearing column from $T^+$ to simulate partial relevance. 
We train $f_\theta$ using the following margin-based ranking loss:
\[
\mathcal{L} = \sum_{(q, T^+, T^-)} \max(0, 1 - f_\theta(q, T^+) + f_\theta(q, T^-)).
\]
At inference time, $f_\theta$ is used to rerank the retrieved candidates based on semantic coverage.

\subsection{Ensuring Completeness using Coverage Verification}
\label{sec:coverage_verification}

Although reranking improves the candidate quality, non-relevant information in the retrieval stage may still lead to only partial coverage of the information needs. 
To address this problem, we introduce a coverage verification function using a gap detection and refinement algorithm.

\noindent\textbf{Connected Table Group Construction}. We construct connected table groups by selecting a single table for each sub-question such that the corresponding clusters form a connected component in $\mathcal{G}$, \revise{ inspired by Steiner-tree-based schema graph \cite{ren2024purple}}.
Each group represents a candidate set of tables that collectively cover the full question. 
Each group is assigned a score by concatenating all of the tables involved and applying $f_\theta$ on the concatenated content and question.

\noindent\textbf{Gap Detection and Refinement.} If the score of the top-ranked group is less than a predefined threshold, we assume the coverage is incomplete. 
Next, we use an LLM to generate a residual sub-question following ~\cite{zhang2025murre}.
We then retrieve and rerank candidate tables using this new sub-question and select an alternative table—connected to the current group—that maximizes the joint coverage score.

We finally assign each table a score based on the best-performing group containing it, and select the top-k tables for answer generation.
This design enables higher-precision retrieval that is effective and scalable for large-scale tables.

\smallskip
\update{\noindent\textbf{Remark.} While \methodqa{} provides initial column-level alignments, this mapping is insufficient for robust table retrieval. 
Our empirical analysis shows that this result leads to inaccurate mappings ($60\%$), due to semantic ambiguities, or incomplete mappings ($40\%$), where tables are missed because no single column strongly matches an information need. 
Therefore, a dedicated retriever is essential to identify a complete and relevant set of tables.
}

%% file: ICDE_MTQA/3_method_reasoner.tex
\section{Sub-question Guided Reasoner} \label{sec_method_rs}
Once the relevant tables are retrieved, a common \textit{answer generation} strategy is to
perform the reasoning over tables directly using a sequence-to-sequence model~\cite{pal2023multitabqa}. 
However, this strategy often performs poorly on numerical questions \revise{because the answer must be computed through explicit calculations (e.g., standard aggregations and arithmetic operations), rather than obtained by simple lookup.}
Program-based reasoning \cite{zhu2024autotqa} offers a structured alternative -- by first generating an executable program (e.g., SQL or Python) that is then executed over multiple tables to produce the final answer.
\revise{However, most existing program generators are designed for standard Text-to-SQL settings over a single database with explicit PK-FK constraints.} When adapted to our setting, namely a collection of \revise{ isolated tables without explicit inter-table relationships}, they often become unreliable: they may select non-relevant tables, fail to infer which tables can be joined, or produce incorrect reasoning steps~\cite{DBLP:journals/corr/abs-2501-09310}.

To address these challenges, we propose \textbf{Sub-question Guided Reasoner}. 
\hlight{Instead of generating the entire program in one shot, our approach incrementally constructs the program using a sequence of decomposed sub-questions. 
Each sub-question can be reliably answered using a single table or a unionable group, and inter-sub-question dependencies determine which intermediate results should be joined to get a complete reasoning program. 
The resulting program is then executed to derive the final answer.
This design leads to a more accurate reasoning pipeline,
capable of supporting diverse types of reasoning programs over isolated retrieved tables.}

\subsection{Program Generation}
\label{sec:program_gen}
Given the original question, a set of decomposed sub-questions, and the retrieved tables, our reasoner generates a program in two stages:

\noindent\textbf{CoT-Guided Multi-step Program Generation.} 
Following recent advances in multi-hop reasoning~\cite{wei2022chain}, we integrate chain-of-thought (CoT) prompting to generate program step by step based on the sequence of sub-questions.
The process begins by generating an initial sub-program that uses a unionable group of tables relevant to the initial sub-question.
Then, for each subsequent sub-question, the program is incrementally improved by joining the current intermediate program with a new sub-program generated for that sub-question.
The final program is obtained after processing all sub-questions.

This step-by-step process provides two benefits: (1) It explicitly encodes reasoning to solve high complexity tasks based on table dependencies; (2) It enables more robust program construction by constructing each reasoning step based on smaller, coherent subsets of the table collection. 
To handle column-level inconsistencies when joining unionable tables, we also include a fuzzy-join operator~\cite{khatiwada2025fuzzy} \revise{for aligned textual join keys.}

\noindent\textbf{Execution-guided Refinement.} Despite the use of carefully constructed prompts, the generated program may still contain errors (e.g., syntax errors). 
To improve robustness, we introduce an \emph{execution-guided refinement} step in our solution. 
We first execute the program using the retrieved tables and check the output for any failures. 
If errors are detected, we re-prompt the LLM and include the error message to help refine the program. 
This process is repeated until a valid program is generated or a maximum retry limit is reached.

\smallskip

%% file: ICDE_MTQA/4_0_dataset_preparation.tex
\section{Data Preparation for Evaluation} \label{sec:data_prep}
In this section, we first examine the limitations of existing benchmarks for our problem setting.
Then, we introduce our designed solutions to prepare the evaluation datasets.

\subsection{Motivation}
\hlight{
To the best of our knowledge, there are no existing benchmarks that directly support the evaluation of numerical MTQA over \settingdef{}.
First, datasets commonly adapted for MTQA evaluation~\cite{chen2024table, wummqa} are directly sourced from text-to-SQL benchmarks, such as Spider~\cite{yu2018spider}, and Bird~\cite{li2023can}.
These original and derived benchmarks typically consist of tables from one or multiple databases, with limited scale, complete metadata, and focus only on explicit PK-FK joins.
Thus, they fail to reflect the \textit{scalability}, \textit{incomplete metadata}, and \textit{complex table relationships} in \settingdef{}.
Second, popular table QA benchmarks, such as NQ-Tables \cite{herzig2021open} and Open-WikiTables \cite{kweon2023open}, are designed for open-domain single QA, where each question can be answered using a single table.
Thus, they lack the multi-table question–answer pairs for evaluation.
}

\subsection{Preparation Process} \label{sec:data_pre_solution}
To address these gaps, our data preparation follows two steps. 
\hlight{
First, we collect real-world question–answer pairs that are explicitly grounded in multiple relevant tables.
Second, we expand these grounded tables into a large-scale table repository, ensuring the \textit{scalability}, \textit{incomplete metadata}, and \textit{complex table relationships} characteristic of \settingdef{}.}
To realize this, we repurpose the question–answer pairs and their grounded tables from existing text-to-SQL benchmarks, Spider~\cite{yu2018spider} and Bird~\cite{li2023can}. 
The resulting datasets are referred to as SpiderWild and BirdWild.
\hlight{
Next, we will introduce how we prepare \settingdef{} by repurposing grounded tables and introducing external tables, and make the question annotation.}

\subsubsection{Repurposing Grounded Tables} 
For each text-to-SQL benchmark, we begin by \revise{ collecting} all tables from the original databases as the ground tables into a centralized repository. To reflect the unique characteristics, we then apply a three-stage table transformation pipeline:

\smallskip
\noindent\textbf{Stage 1: Table Decomposition.} To overcome the limited scale and inter-table relationship diversity of the original datasets, we adopt the idea of table decomposition inspired by \cite{deng2024lakebench}.
Different from their random decomposition, our decomposition is designed to produce more semantically meaningful tables.
Specifically, we decompose large tables (i.e., those with more than 5 columns and 50 rows) into multiple disjoint subtables using both column-wise and row-wise splitting strategies.
This design is motivated by common organizational patterns observed in data lakes, where tables are often constructed using semantically related columns (e.g., \texttt{Year}, \texttt{Month}, \texttt{Day}) or partitioned by value ranges or categories \cite{wang2025www, nargesian2020organizing}.

For the column-wise splitting strategy, we first identify key columns (i.e., columns with all distinct values) and use the LLM to group the remaining non-key columns into semantically related column subsets.
To achieve this, we instruct the LLM to cluster columns by topic or theme based on their column headers.
Each column group is then combined with a randomly selected key column to form a new subtable.
To ensure the decomposed subtables can be joined together to reconstruct the original table (i.e., values from the same row in the original table remain correctly aligned), we verify whether any subtable contains all key columns. If no such subtable exists, a subtable containing all key columns is created.
Since the original database tables only provide the PK-FK joins, we aim to introduce additional joinability of semantic joins. Thus, we choose subtables derived from the same key column that does not participate in PK-FK joins to create joinability. 

Row-wise splitting strategy is applied after the column-wise step, which partitions each subtable based on the distribution of a randomly selected non-key column. 
If a numerical column is chosen, we partition the rows into a random number of buckets (between 5 and 20) based on value ranges (e.g., splitting a \texttt{sales} table by ranges of \texttt{sale amounts}).
If a categorical column is chosen, we randomly divide the table into 2 to 20 tables, each containing rows for a specific group of categories (e.g., splitting a \texttt{sales} table by \texttt{region} might yield separate tables for ``North America Sales'', ``Europe Sales'' etc.). 
This row-wise splitting produces unionable tables that share the same headers but cover different table content, \update{which mirrors the common real-world occurrence of unionable tables, as seen in the data versioning of Wikipedia tables \cite{bleifuss2021vldb}.}
\revise{After decomposition, we manually inspect the subtables to verify that they preserve the original semantics (e.g., the original table can be reconstructed by joining/unioning subtables).}

\smallskip
\noindent\textbf{Stage 2: Metadata Incompleteness Simulation.} To mimic the incomplete metadata, we randomly select 20\% of the decomposed tables and mask 50\% of their column headers and table titles, using the placeholder \texttt{MASK} based on the incomplete metadata statistics reported in~\cite{khatiwada2023santos,solo2023sigmod}. 
For example, an \texttt{employee} table with columns \texttt{emp\_id}, \texttt{department}, \texttt{salary}, and \texttt{role} may be transformed into \texttt{MASK} table with columns \texttt{MASK}, \texttt{department}, \texttt{MASK}, and \texttt{role}.

\noindent\textbf{Stage 3: Joinability Simulation.} Among various types of joinability, fuzzy joins (where values differ slightly in spelling or format) are particularly common in real-world data lakes \cite{zhu2017auto, ji2025vldb, hu2025polyjoin}. To simulate this, we inject value-level variations into the textual key columns of decomposed tables. Specifically, following \cite{hu2025polyjoin}, we introduce perturbations such as typographical errors and character deletions into 20\% of the cell values. This transformation creates realistic join scenarios where approximate string matching is required (e.g., \texttt{"New York"} vs. \texttt{"Nw York"}).

\subsubsection{Incorporating External Tables}  \label{sec_dataprep_external_tabs}
While our table decomposition strategy increases the number of tables from hundreds to thousands, it remains insufficient to reflect the large-scale settings typically encountered in practice, such as the massive collections of web tables available online. To further enhance scalability, we incorporate additional tables from the widely used table corpora WebTables and OpenData ~\cite{deng2024lakebench,luo2026vldb} \revise{ considering two dimensions:  
(1) tables semantically related to the questions—as the hard-negatives, and  
(2) tables from a similar domain as existing tables, where domain relevance is estimated using the relevance of table titles--to simulate realistic distractor tables.}
Specifically, for each collected question or decomposed table, we retrieve the top-$N$ candidate tables from these corpora using BM25 \cite{robertson2009ftir} over both the question and the table names. 
This enrichment significantly enlarges the table repository. For example, with $N{=}100$, the number of tables increases from 2,210 to 73,688 on SpiderWild, and from 5,136 to 109,949 on BirdWild.

\subsubsection{Question Annotation} 
Finally, we curate a set of numerical questions from the original questions that require reasoning over multiple tables. For each selected question, we then construct annotations consisting of both the ground-truth answer and the set of relevant tables to answer this question.
Specifically, we first execute the original SQL query on their provided databases to obtain the ground-truth answer.
\revise{ Then we write and execute an adapted SQL query over the decomposed tables (via join/union operators) to verify the same answer.}
Next, to determine the relevant tables, we analyze how the original SQL query draws information from multiple tables. 
\hlight{ To achieve this, we manually rewrite the query to extract the values of the primary key and other relevant columns (e.g., those mentioned in WHERE or GROUP BY clauses) for each relevant table.
We then identify the specific table rows and columns containing these values.
These records are then mapped back to the decomposed tables in the table repository by locating the tables that contain the corresponding rows and columns. 
The decomposed table subset that jointly covers all these records is regarded as the relevant tables.}
\revise{ We also spot-check 10\% of questions to verify that no alternative feasible solutions are introduced from external tables.}

%% file: ICDE_MTQA/4_experiment.tex
\section{Experimental Evaluation} \label{sec_exp_total}
\noindent\textbf{Evaluation Goals.}  
\input{ICDE_MTQA/tables/experiments/data_statistics}
We aim to assess the effectiveness, efficiency, scalability, and robustness of our proposed method in realistic settings of MTQA over \settingdef{}, and provide a secondary comparison with Text-to-SQL and Open-domain single-table QA methods. Specifically, we answer the following five questions:

\input{ICDE_MTQA/tables/experiments/main_res_retrieve_effectiveness}

\begin{itemize}[leftmargin=*,noitemsep]
    \item \textbf{Q1: How well does our method retrieve multiple relevant tables and answer numerical questions?}
    Unlike prior work in open-domain MTQA, which struggles to retrieve relevant tables at scale and generate reliable answers (\S \ref{sec_related_work}),
    our framework aims to address such limitations. We evaluate the effectiveness and efficiency, in both table retrieval and answer generation, against strong baselines.  
    (\S~\ref{sec_exp_main_res})

    \item \textbf{Q2: What is the contribution of our key design choices and parameter settings to overall performance?} 
    We perform ablation studies to assess the impact of core components (i.e., preprocessing, decomposer, retriever, and reasoner) and analyze the sensitivity of key parameters such as table joinability and unionability thresholds.  
    (\S~\ref{sec_exp_ablation})

    \item \textbf{Q3: How well does our method scale with the size of the table corpus?}  
    We evaluate whether \ourmethod{} maintains high effectiveness and low latency as the number of tables in the table corpora increases.  
    (\S~\ref{sec_exp_scale})

    \item \textbf{Q4: How robust is our method under different challenging scenarios?} 
    In particular, we evaluate its robustness w.r.t. the varying number of involved tables, their unionability, and degrees of completeness of metadata.  
    (\S~\ref{sec_exp_robustness})

    \item \textbf{Q5: How does our method compare with Text-to-SQL and Open-domain single-table QA methods?} 
    We conduct the evaluation against Text-to-SQL (\S~\ref{sec_exp_text2sql_01}) and open-domain single-table QA methods (\S~\ref{sec_exp_tqa}). 
\end{itemize}

\subsection{Evaluation on Table Retrieval and Answer Generation} \label{sec_exp_main_res}
\subsubsection{Experimental Setup}  \label{sec_exp_setup}
We evaluate \ourmethod{} on the prepared SpiderWild and BirdWild datasets.
There are three primary factors that directly impact MTQA: (1) the number of table joins, (2) the presence of union operations, and (3) whether the relevant tables contain incomplete metadata. 
To facilitate a more fine-grained evaluation over these factors, we categorize the questions in both datasets into three levels of complexity—\textit{Easy}, \textit{Moderate}, and \textit{Hard}—based on the number of table joins and unions required to derive the answer, and whether incomplete metadata is involved.
Specifically, \textit{Easy} questions involve two table joins, require no union, and rely on complete metadata.  
By contrast, \textit{Hard} questions require more than two joins, at least one union, and involve incomplete metadata in the relevant tables.
Questions that do not meet the criteria for Easy or Hard category are classified as \textit{Moderate}.
Table~\ref{tab:dataset_statistics} summarizes the dataset statistics.

\noindent\textbf{Evaluation Metrics.}  
Following previous work ~\cite{chen2024table,wummqa}, we use Precision@k (P@k), Recall@k (R@k), and F1@k to evaluate the effectiveness for multi-table retrieval. 
To measure answer accuracy on numerical questions, we adopt Arithmetic Exact Match (EM@k) from \cite{wang2024enhancing}, which measures the percentage of answers obtained by executing the generated program using top-k retrieved tables that match the ground truth.
We choose 3 and 5 for k following the previous study ~\cite{wummqa}.

\noindent\textbf{Implementation \& Hardware.}
We use the BGE-M3 \cite{bgem3} for embedding, due to its effectiveness in retrieval tasks~\cite{zhu2025mitigating}. 
We set the unionability and joinability thresholds to 0.9 and 0.5, respectively.
We include \texttt{GPT-4.1 mini} as the primary LLM model. 
The prompts used by our framework are provided in 
\ifcameraready
\else
Appendix \ref{prompt_supp}.
\fi
All experiments were conducted on a server running Red Hat 7.9, equipped with an Intel(R) Xeon(R) E5-2690 CPU, 512GB RAM, and a 16GB NVIDIA Tesla P100 GPU. Our source code and benchmarks are available at \cite{tr}.

\noindent\textbf{Competitors.}
We first evaluate our approach against JARUnion and MMQAUnion, which are adapted from the state-of-the-art open-domain MTQA systems JAR~\cite{chen2024table} and MMQA~\cite{wummqa}, respectively.
Both methods were originally designed for databases with complete metadata and do not consider table unionability.
To ensure a fair comparison, we apply their original table retrieval strategies on our processed metadata-complete tables and augment the retrieved tables by incorporating unionability
\ifcameraready
\else
(details are in Appendix~\ref{exp_baselines}).
\fi

\input{ICDE_MTQA/tables/experiments/main_res_reasoning_effectiveness}

\subsubsection{Effectiveness for Multi-table Retrieval}
Table~\ref{baseline_comparison_prf1} presents the table retrieval effectiveness results across varying levels of question complexity. 
We highlight two key observations:  
(1) \ourmethod{} consistently outperforms all baselines across both datasets and all complexity levels.
This gain mainly comes from our decomposition-driven multi-table retrieval strategy: the decomposer produces higher-quality sub-questions, and the retriever then selects a coherent set of tables by explicitly maximizing question coverage. 
While JARUnion and MMQAUnion lack an effective multi-table retrieval mechanism.
(2) Almost all methods exhibit a noticeable performance drop as question complexity increases.  
These trends highlight an increased difficulty of retrieving relevant tables when questions require more joinable and unionable tables, especially in the presence of incomplete metadata.

\subsubsection{Effectiveness for Answer Generation}
Table~\ref{baseline_comparison_em} presents the answer accuracy using the top-k retrieved results. 
We report EM scores under both Top-3 and Top-5 retrieval settings, stratified by question complexity.
We make several key observations:  
(1) \ourmethod{} consistently outperforms all baselines across both datasets and all complexity levels, driven by more accurate table retrieval and program generation.
(2) As question complexity increases, answer accuracy also drops significantly on both datasets.  
This suggests that more complex questions—those involving more joins, unions, and incomplete metadata present greater challenges not only in identifying the correct tables but also in generating correct answers via reasoning.
(3) Interestingly, the drop in answer accuracy across complexity levels (e.g., 70\% of EM@3 on BirdWild from Easy to Hard)  is much larger than that of multi-table retrieval (e.g., 2\% of R@3). 
This indicates that even with accurate table retrieval, reasoning over multiple tables becomes significantly more challenging as the complexity of joins, unions, and incomplete metadata increases.
A systematic error analysis of table retrieval and answer generation is
\ifcameraready
\else
in Appendix \ref{exp_error_analysis}.
\fi

\input{ICDE_MTQA/tables/experiments/main_res_retrieve_reason_timecost}

\subsubsection{Efficiency for Multi-table Retrieval and Answer Generation}
To evaluate the efficiency of MTQA methods, we measure the average runtime per question for both multi-table retrieval and answer generation across all questions in each dataset. 
Note that for the runtime of multi-table retrieval of \ourmethod{}, we include the time consumed by both the question decomposer and retriever modules.
Figure~\ref{baseline_tradeoff_analysis} presents a comparative analysis of efficiency, along with the corresponding effectiveness.
From the results, we observe:
(1) For table retrieval, \ourmethod{} achieves a strong balance between effectiveness and latency. 
Compared to the most efficient baseline, MMQAUnion, it yields an average of 15\% higher R@5 on both datasets, incurring only a $1.6\times$ increase in runtime.
(2) For answer generation, \ourmethod{} delivers substantial accuracy gains—achieving 18\% and 31\% higher EM@5 scores on SpiderWild and BirdWild, respectively, compared to the second-best method JARUnion, at the cost of an average 3.7 seconds per question.
This additional time cost stems from two reasoning-oriented mechanisms incorporated in the reasoning component of \ourmethod{} to enhance program robustness: (i) a step-by-step CoT prompting strategy to generate the reasoning program incrementally, and (ii) a refinement mechanism that verifies and improves the generated program before execution.



\subsection{Ablation Study and Parameter Study } \label{sec_exp_ablation}
\subsubsection{Ablation Study}
We conduct a comprehensive ablation study to evaluate the impact of our key designs -- namely the preprocessing (\S \ref{sec_problem_solution_overview}), decomposer (\S \ref{sec_method_qd}), retriever (\S \ref{sec_method_rt}), and reasoner (\S \ref{sec_method_rs}) components, for the effectiveness.

\input{ICDE_MTQA/tables/experiments/preprocessing_ablation}

\noindent\textbf{Preprocessing.} 
\update{Since the real-world table corpora may contain the incomplete metadata, which can distort unionability calculation during graph construction, we also introduce a metadata inference module. It leverages LLMs to infer missing column headers by incorporating intra-table context, inspired by \cite{sun2023reca} }
\ifcameraready
\else
(details are in Appendix~\ref{meta_infer}).
\fi

In the following, we will evaluate the inference quality of this module against a naive baseline, \textit{Direct LLM}, which directly prompts an LLM to complete missing metadata without using the intra-table context.
Following prior work~\cite{zhang2023nameguess}, we use \textit{BERT-F1} score to assess quality: for each missing column header, we compute the F1 score between the predicted and ground-truth header based on contextual embedding similarity, and report the average across all the missing headers.
Our approach consistently outperforms the \textit{Direct LLM}: on SpiderWild, BERT-F1 increases from 0.602 to 0.697, and on BirdWild from 0.602 to 0.652. These gains indicate that exploiting intra-table context materially improves metadata inference accuracy.

We further evaluate how metadata quality affects downstream retrieval and answer accuracy by comparing four configurations:
(1) \textit{No-fill}, which leaves incomplete metadata unchanged;
(2) \textit{Direct LLM};
(3) \ourmethod{}, which is our proposed metadata inference module;
(4) \textit{GT}, which uses the corresponding ground-truth metadata to replace all incomplete metadata.
The results are presented in Table~\ref{tab:retrieval_answering_fill}, covering both Top-3 and Top-5 retrieval settings.
From the results, we observe that \ourmethod{} consistently improves over the \textit{No-fill} and \textit{Direct LLM} baselines across all evaluation metrics on both SpiderWild and BirdWild datasets. 
These results highlight the importance of completing missing metadata and the effectiveness of using intra-table context for metadata recovery.
Moreover, the performance gap between \ourmethod{} and \textit{GT} setting is relatively narrow, as compared to \textit{No-fill}.
This demonstrates that our inference strategy closely approximates gold metadata and significantly contributes to both table retrieval and answer generation in MTQA.

\noindent\textbf{\methodqa{}.} 
We first compare our decomposition strategy against a \textit{Direct LLM} generation approach, which uses an LLM to generate the sub-questions without any structural alignment information. 
To assess the quality of decomposition, we use three metrics corresponding to the criteria defined in \S \ref{sec_method_qd} --
(i) \textit{Information Retention Rate (IRR)} to measure completeness: The proportion of questions whose decompositions successfully preserve all information needs required;
(ii) \textit{Subquestion Redundancy (SR)} to measure redundancy: The average pairwise semantic similarity computed  using embeddings generated from a pretrained Sentence-BERT model;
(iii) \textit{Subquestion-Table Alignment Rate (SAR)} to measure table-specificity: Determine if the number of sub-questions matches the number of relevant tables required to answer the question.

\input{ICDE_MTQA/tables/experiments/decomposition_metrics}

From Table~\ref{tab:decomposition_quality}, we observe that our table-aligned question decomposer consistently improves all three metrics using both datasets.
These results demonstrate the effectiveness of our decomposition strategy which leverages the structure of the table corpus.
An additional case study illustrating the improvements is shown in 
\ifcameraready
\else
Appendix \ref{exp_supp}.
\fi


\input{ICDE_MTQA/tables/experiments/decomposition_performance}

Next, we deepen our study of the impact of decomposition
on both retrieval effectiveness and final answer accuracy in Table~\ref{tab:retrieval_and_answering}.
Observe that our decomposition consistently achieves better performance on Top-3 and Top-5 retrieval settings.
These results demonstrate that \ourmethod{} improves both retrieval effectiveness and final answer accuracy by improving the quality of decomposed sub-questions.

\noindent\textbf{\methodrt{}.}
Now, we evaluate the retrieval effectiveness of the coverage scoring function and coverage verification submodule using two ablations:
(1) \textit{NaiveScoring} is used to replace our trained coverage scoring function using a simple baseline which uses the sum of the individual embedding similarities computed between sub-questions and candidate tables, obtained using our coarse retrieval method described previously.
(2) \textit{w/o Verification} disables our residual sub-question generation that is used to retrieve complementary tables.
In Table \ref{tab:coverage_aware_ablation}, we find: (1) Our trained coverage scoring function provides more effective retrieval performance by identifying subsets of tables that cover the question intent more fully.
(2) Our coverage verification submodule enhances retrieval performance by filling potential coverage gaps introduced by non-relevant tables --those that appear to be relevant to the question but cannot provide a
complete answer.

\input{ICDE_MTQA/tables/experiments/retriever_ablation}

\noindent\textbf{\methodrs{}.}
Assessing the benefit of our reasoning module is achieved by comparing it to two other variants:
(1) A \textit{w/o CoT} baseline that prompts the LLM to generate a program in a single shot using the original question and the most similar tables retrieved; 
(2) A \textit{w/o Refinement} executes the first program generated with no further refinement. 
The answer accuracy is shown in Table~\ref{tab:reasoning_ablation}, which illustrates that: (1) Our chain-of-thought generation substantially improves answer quality, achieving up to 19\% better EM scores. 
This result demonstrates the benefits of modeling the reasoning process using sub-questions, which enables the LLM to correctly infer table relationships for generating an accurate program, thus producing reliable answers.
(2) Our execution-guided refinement further improves answer quality by ensuring the correctness of the generated program.

\input{ICDE_MTQA/tables/experiments/reasoner_ablation}

\input{ICDE_MTQA/tables/experiments/impact_params}
\subsubsection{Parameter Study} \label{sec_exp_res_param}
We conduct a parameter study to investigate the impact of different thresholds for \textit{joinability} and \textit{unionability} when modeling table relationships (\S~\ref{sec_problem_solution_overview}). 
We set the threshold ranges—[0.1, 0.3, 0.5, 0.7, 0.9] for joinability and [0.5, 0.6, 0.7, 0.8, 0.9] for unionability. 
The lower bounds are selected based on their strong pruning effects observed across all table pairs, where they effectively filter out a large portion of low-quality or spurious relationships across tables.
Figure~\ref{fig:impact_scale_test_compact} illustrates the retrieval and answering effectiveness under varying threshold settings.
Observe that:  
(1) For joinability, both performances improve as the threshold increases from 0.1 to 0.5, then decline.
This is because very low thresholds (e.g., 0.1) admit many noisy or spurious links between unrelated tables, which dilute the quality of retrieval and introduce irrelevant candidates for reasoning. While very high thresholds (e.g., 0.9) are overly restrictive, they exclude moderately joinable tables, which are still necessary to derive the correct answer.
(2) For unionability, both performance consistently improves as the threshold increases. 
This is because stricter thresholds ensure that only highly unionable tables—those with strong semantic similarity—are grouped together. This reduces the inclusion of semantically misaligned tables in union groups, thereby enhancing both retrieval precision and the grounding quality for reasoning.

\subsection{Scalability Study} \label{sec_exp_scale}
\subsubsection{Experimental Setup} To evaluate the scalability of \ourmethod{} under varying sizes of table corpus, we vary the number of top-$N$ external tables retrieved per question/table (\S~\ref{sec_dataprep_external_tabs}) with $N \in \{100, 150, 200, 250, 300\}$. 
These settings yield table corpora containing approximately 109K, 149K, 183K, 214K, and 243K tables, respectively.
Since the scale of the table corpus largely influences the table retrieval effectiveness and efficiency, we are mainly focusing on the retrieval effectiveness measured by R@5, and the retrieval efficiency measured by the retrieval time cost per question.

\input{ICDE_MTQA/tables/experiments/scale_test}
\subsubsection{Main results} 
Figure~\ref{fig:impact_scale_test_compact} presents the results. As the size of the table corpus increases, we observe that \ourmethod{} remains robust in retrieval effectiveness—exhibiting only a modest 5\% drop in R@5 despite more than doubling the number of tables. In terms of efficiency, the query time grows gradually with scale, indicating that the method remains computationally efficient even under a larger table corpus.
\update{For the offline graph construction efficiency, using the largest 243K tables on both datasets requires approximately 4.2 days to compute joinability (with 5 parallel workers), while unionability takes about one day since it operates over table pairs which are fewer than the column pairs required for joinability.
\hlight{
Looking ahead, we plan to reduce this preprocessing cost via distributed engineering techniques (e.g., sharding and parallel similarity joins) \cite{efficientcomputation}. 
We also plan to support scalable, low-cost dynamic updates so that new tables and relationships can be incorporated incrementally without full recomputation.
}
}

\subsection{Robustness Cross Challenge Scenarios} \label{sec_exp_robustness}
\noindent\textbf{Robustness on Questions Requiring a Varying Number of Relevant Tables.}
To evaluate robustness across questions involving different numbers of tables, we group questions by the number of relevant tables retrieved (2 vs. $\geq$3) and assess both retrieval and answer effectiveness.
As shown in Figure~\ref{tab:reltabcnt_impact}, \ourmethod{} consistently outperforms all other baselines for both groupings, demonstrating the robustness of our solution on questions of higher complexity.
This is because: (1) Our retriever explicitly optimizes for question coverage, ensuring that the necessary tables are selected even as the required set grows; (2) Our reasoner generates programs that are guided using decomposed sub-questions, which enables accurate inference of table relationships under increased difficulty.
\input{ICDE_MTQA/tables/experiments/impact_reltabcnt}

\input{ICDE_MTQA/tables/experiments/impact_metadata}
\input{ICDE_MTQA/tables/experiments/impact_union}


\noindent\textbf{Robustness of Questions Involving Incomplete Metadata.} 
We also consider the robustness of our method by comparing the performance on questions with relevant tables, which include incomplete metadata, with the complete metadata case.
The results are shown in Figure \ref{tab:metadata_impact}.
Observe that: (1) All methods have degraded performance when introducing questions that require tables that have incomplete metadata.
This highlights the challenges that incomplete metadata can introduce. 
(2) \ourmethod{} is consistently the most effective approach for retrieval and answer effectiveness across both datasets, with notably less performance degradation when compared to the baselines (e.g., an average EM@5 drop of 26\% for \ourmethod{} vs. 62\% for the baselines).
These results demonstrate \ourmethod{} is more robust when incomplete metadata exists.

\noindent\textbf{Robustness on Questions that Require Unionable Tables.} We now compare the performance of our method using questions that require unionable tables to answer versus those that do not. The results are shown in Figure~\ref{tab:union_impact}.
Once again, \ourmethod{} is consistently the most effective retrieval model and produces higher-quality answers for both question types and datasets.
While existing baselines have notable performance degradation when answering questions that require one or more union operations, \ourmethod{} is consistently good in both scenarios. 
The advantage stems from our design decisions, which explicitly model table unionability using a graph and groups unionable tables into clusters during the retrieval and reasoning stage. 
In contrast, the extended baselines assume that each table is independent and fail to incorporate unionability into the retrieval model, reducing their ability to locate multiple joinable tables and use them for reasoning.


\subsection{Comparison with Text-to-SQL methods}\label{sec_exp_text2sql_01}
\hlight{Text-to-SQL methods are designed for databases with complete metadata and explicit table relationships. This setting provides advantages that are not present in our scenario. 
To provide a fair comparison, we conduct two complementary evaluations: (1) an adapted evaluation that runs Text-to-SQL on the same collections, SpiderWild and BirdWild, via database materialization (\S \ref{sec_exp_text2sql_01_01}), and (2) a secondary evaluation under their original settings, where Text-to-SQL methods are given full database schema while \ourmethod{} is provided only with the table collection collected from the corresponding relational database for each question (\S \ref{sec_exp_text2sql_01_02}).}

\revise{
\subsubsection{Adapted Evaluation on SpiderWild and BirdWild} \label{sec_exp_text2sql_01_01}
To enable a fair comparison on the large-scale table collections SpiderWild and BirdWild, we compare \ourmethod{} with an adapted Text-to-SQL baseline, OpenSearch-Wild, based on the high-performing OpenSearch-SQL~\cite{xie2025opensearchsqlenhancingtexttosqldynamic}, as reported in \cite{birdsqlleader}. Since OpenSearch-SQL operates over a single database and does not include a table-retrieval stage, we adapt it to our setting by adding (i) database materialization during its Preprocessing based on our inferred metadata and Table Relationship Graph, and (ii) top-k table selection during its Extraction based on its obtained columns 
\ifcameraready
\else
(details are in Appendix~\ref{exp_baselines}).
\fi

\noindent\textbf{Main Results.}
We summarize the effectiveness, efficiency, and robustness of the adapted baseline OpenSearch-Wild and \ourmethod{} as follows.
(1) As shown in Table~\ref{baseline_comparison_prf1}, \ourmethod{} consistently outperforms OpenSearch-Wild. This is because OpenSearch-Wild relies on question-column relevance and is sensitive to column-mapping errors, while \ourmethod{} constructs coherent table sets by explicitly maximizing question coverage, thereby reducing cascading retrieval errors.
(2) As shown in Table~\ref{baseline_comparison_em}, OpenSearch-Wild still lags behind \ourmethod{}. This is mainly because its weaker table identification often misses required tables for SQL generation.
(3) As shown in Figure~\ref{baseline_tradeoff_analysis}, OpenSearch-Wild incurs the highest latency in both retrieval and answer generation. For retrieval, it adds an online column-filtering stage that selects columns whose semantic similarity to the question exceeds a threshold, requiring embedding computation and similarity scoring. For answer generation, it uses multi-round, step-wise prompting to derive SQL elements (e.g., SELECT clause and values) before producing the final SQL.
(4) Across varying numbers of involved tables, unionability, and metadata completeness (Figures~\ref{tab:reltabcnt_impact}--\ref{tab:union_impact}), \ourmethod{} exhibits more stable performance, due to its robust multi-table retrieval and reasoning framework.
}

\subsubsection{Secondary Evaluation using Original Settings} \label{sec_exp_text2sql_01_02}
\revise{
We evaluate on three widely used Text-to-SQL benchmarks—Spider~\cite{yu2018spider}, BIRD-dev~\cite{li2023can}, and ScienceBenchmark~\cite{zhang2023sciencebenchmark}. 
From each benchmark, we focus on the curated numerical subset whose questions require reasoning over multiple tables. 
In total, Spider includes 274 numerical questions over 79 databases with 437 tables; BIRD-dev includes 45 numerical questions over 7 databases with 52 tables; and ScienceBenchmark includes 36 numerical questions over 3 databases with 50 tables.}
We compare against two resource-efficient, high-performing Text-to-SQL systems CHESS and OpenSearch-SQL (\S \ref{sec_related_work}), as reported in \cite{birdsqlleader}.
We report two metrics: table selection accuracy, which computes the percentage of questions whose predicted SQL references exactly the ground-truth relevant tables;
and answer accuracy, which corresponds to the execution accuracy \cite{yu2018spider} in Text-to-SQL.


\revise{
\noindent\textbf{Main Results.}
As shown in Table~\ref{tab:text2sql_results}, \ourmethod{} achieves strong table identification: it performs best on Spider and ScienceBenchmark and remains competitive on Bird-dev compared to the best-performing OpenSearch-SQL. This indicates that \ourmethod{} can effectively identify the correct table subset despite lacking access to PK-FK constraints.
In terms of answer accuracy, \ourmethod{} performs best on Spider and ScienceBenchmark. We attribute these gains to its stronger table identification, which provides the reasoner with more complete relevant tables. On Bird-dev, we observe a gap compared to OpenSearch-SQL. 
 Our error analysis shows that most failures (57\%) stem from value grounding, i.e., values in the generated SQL do not match column values. 
This is expected because our focus is robust multi-table retrieval and answering for a large scale of tables, rather than optimizing SQL generation.
Our reasoner does not include a value-mapping module due to the substantial value preprocessing and storage overhead, while OpenSearch-SQL incorporates value-level extraction for value grounding. 
In future work, we will enhance our reasoner with lightweight value mapping to improve SQL correctness.}

\input{ICDE_MTQA/tables/experiments/text2sql_comparison}

\revise{
\subsection{Comparison with Open-domain Single-table QA methods} \label{sec_exp_tqa}
For this comparison, we use the same three datasets as \ourmethod{} (\S \ref{sec_exp_text2sql_01_02}).
We compare against DTR~\cite{herzig2021open}, a dense table retriever paired with an LLM-based programmatic reasoner~\cite{chen2024table}, and TableRAG~\cite{yu2025tablerag}, an iterative retrieval-augmented generation approach that conditions an LLM on a retrieved table set to produce the answer.
We report table selection accuracy, defined as the fraction of questions whose retrieved table set exactly matches the gold relevant tables, and answer accuracy, defined as the fraction of questions whose predicted answer matches the gold answer.

\noindent\textbf{Main Results.}
Table~\ref{tab:text2sql_results} shows that \ourmethod{} substantially outperforms both single-table QA baselines in table identification and answer accuracy, highlighting the importance of modeling inter-table relationships for MTQA.

}

%% file: ICDE_MTQA/tables/experiments/data_statistics.tex
\begin{table}[t]
\centering
\small
\setlength{\tabcolsep}{5pt}
\begin{threeparttable}
\begin{tabular}{lcc}
\toprule
Property & SpiderWild& BirdWild\\
\midrule
\# Tables & 73{,}688& 109{,}949 \\
\% Joinability & 9\% & 13\% \\
Avg. \# Columns & 4.4& 5.0\\
Avg. \# Rows & 1{,}384& 8{,}629\\
\midrule
\# Numerical Questions & 274 & 461 \\
 \% Easy/Moderate/Hard& 52\%/43\%/5\% & 10\%/84\%/6\%\\
\% Relevant Table (2~/~$>=$3)
  & 67\%  / 33\%  & 64\%  / 36\%  \\
\% Incomplete Metadata & 42\%  & 48\%  \\
\% Requiring Unionability &   13\%  &  56\%  \\
\bottomrule
\end{tabular}
\captionsetup{skip=0pt}
\caption{Dataset statistics.}
\label{tab:dataset_statistics}
\end{threeparttable}
\vspace{-3mm}
\end{table}

%% file: ICDE_MTQA/tables/experiments/main_res_retrieve_effectiveness.tex

\begin{table*}[t]
\centering
\small
\setlength{\tabcolsep}{3.5pt}
\definecolor{best}{gray}{0.9}
\begin{tabular}{llccc|ccc|ccc|ccc|ccc|ccc}
\toprule
\multirow{3}{*}{Top-k} & \multirow{3}{*}{Method} & \multicolumn{9}{c}{SpiderWild} & \multicolumn{9}{c}{BirdWild} \\
\cmidrule(lr){3-11} \cmidrule(lr){12-20}
&& \multicolumn{3}{c|}{Easy} & \multicolumn{3}{c|}{Moderate} & \multicolumn{3}{c|}{Hard} 
   & \multicolumn{3}{c|}{Easy} & \multicolumn{3}{c|}{Moderate} & \multicolumn{3}{c}{Hard} \\
&& P& R& F1& P& R& F1& P& R& F1& P& R& F1& P& R& F1& P& R& F1\\
\midrule
\multirow{3}{*}{Top-3}
& JARUnion     & 46.5 & 69.7 & 55.8 & 45.2 & 60.6 & 51.3 & 44.8 & 59.8 & 50.7 & 39.7 & 55.0 & 45.8 & 39.2 & 53.7 & 45.7 & 45.0 & 53.7 & 44.8 \\
& MMQAUnion    & 47.2 & 70.8 & 65.6 & 37.9 & 52.4 & 43.7 & 37.9 & 52.1 & 43.5 & 41.0 & 56.6 & 49.1 & 36.7 & 49.7 & 41.8 & 35.9 & 48.8 & 41.0 \\
& OpenSearch-Wild    & 43.9 & 65.8 & 52.7 & 42.4 & 50.1 & 45.4 & 36.7 & 45.0 & 40.0 & 49.5 & 67.4 & 56.5 & 40.4 & 60.6 & 48.5 & 36.8 & 43.0 & 39.0 \\

& \ourmethod{} & \textbf{54.9} & \textbf{82.4} & \textbf{65.9} 
               & \textbf{52.9} & \textbf{71.3} & \textbf{60.2}
               & \textbf{52.3} & \textbf{70.3} & \textbf{59.4}
               & \textbf{52.2} & \textbf{71.0} & \textbf{59.6} & \textbf{52.7}
               & \textbf{70.7} & \textbf{59.8} & \textbf{51.5}
               & \textbf{69.5} & \textbf{58.6} \\
\midrule
\multirow{3}{*}{Top-5}
& JARUnion     & 31.3 & 78.2 & 44.7 & 32.3 & 71.5 & 44.1 & 31.9 & 70.4 & 43.5 & 28.6 & 65.6 & 39.5 & 28.4 & 64.4 & 39.1 & 28.1 & 64.1 & 38.8 \\
& MMQAUnion    & 31.1 & 77.8 & 44.5 & 26.4 & 60.3 & 36.5 & 26.2 & 59.7 & 36.2 & 26.7 & 60.9 & 36.9 & 26.9 & 60.7 & 37.0 & 26.3 & 59.6 & 36.2 \\
& OpenSearch-Wild    & 30.6 & 76.4 & 43.7 & 28.3 & 65.2 & 39.5 & 24.0 & 50.8 & 32.6 & 32.1 & 72.8 & 44.2 & 24.8 & 62.1 & 35.5 & 29.0 & 55.0 & 37.3 \\
& \ourmethod{} & \textbf{34.2} & \textbf{85.6} & \textbf{48.9} 
               & \textbf{33.1} & \textbf{74.4} & \textbf{45.4}
               & \textbf{32.7} & \textbf{73.5} & \textbf{45.0}
               & \textbf{33.3} & \textbf{74.8} & \textbf{45.4}
               & \textbf{33.3} & \textbf{74.7} & \textbf{45.7}
               & \textbf{32.8} & \textbf{73.7} & \textbf{45.0} \\
\bottomrule
\end{tabular}
\captionsetup{skip=0pt}
\caption{Comparison of table retrieval effectiveness on SpiderWild and BirdWild datasets.}
\vspace{-3mm}
\label{baseline_comparison_prf1}
\end{table*}

%% file: ICDE_MTQA/tables/experiments/main_res_reasoning_effectiveness.tex

\begin{table}[t]
\centering
\small
\setlength{\tabcolsep}{1pt}
\begin{tabular}{llccc|ccc}
\toprule
\multirow{2}{*}{Top-k} & \multirow{2}{*}{Method} 
& \multicolumn{3}{c}{SpiderWild} 
& \multicolumn{3}{c}{BirdWild} \\
\cmidrule(lr){3-5} \cmidrule(lr){6-8}
& & Easy & Moderate & Hard & Easy & Moderate & Hard \\
\midrule
\multirow{3}{*}{Top-3}
& JARUnion     & 40.8 & 28.9 & 10.0 & 28.8 & 24.9 & 1.2 \\
& MMQAUnion    & 40.8 & 25.6 & 14.0 & 26.7 & 24.9 & 3.8 \\
& OpenSearch-Wild    & 47.8 & 42.1 & 20.0 & 41.6 & 41.8 &20.0 \\
& \ourmethod{} & \textbf{57.0} & \textbf{43.8} & \textbf{20.0} & \textbf{45.7} & \textbf{43.3} & \textbf{26.9} \\
\midrule
\multirow{3}{*}{Top-5}
& JARUnion     & 46.5 & 38.0 & 30.0 & 41.7 & 36.0 & 3.8 \\
& MMQAUnion    & 46.5 & 33.9 & 28.0 & 35.1 & 31.9 & 11.5 \\
& OpenSearch-Wild    & 47.0 & 40.5 & 20.0 & 40.7 & 39.8 & 20.0 \\
& \ourmethod{} & \textbf{57.7} & \textbf{41.3} & \textbf{35.0} & \textbf{47.5} & \textbf{44.6} & \textbf{19.2} \\
\bottomrule
\end{tabular}
\captionsetup{skip=0pt}
\caption{Answer accuracy comparison over retrieved tables.}
\vspace{-3mm}
\label{baseline_comparison_em}
\end{table}

%% file: ICDE_MTQA/tables/experiments/main_res_retrieve_reason_timecost.tex
\begin{figure}[t]
  \centering
  \includegraphics[width=0.8\linewidth]{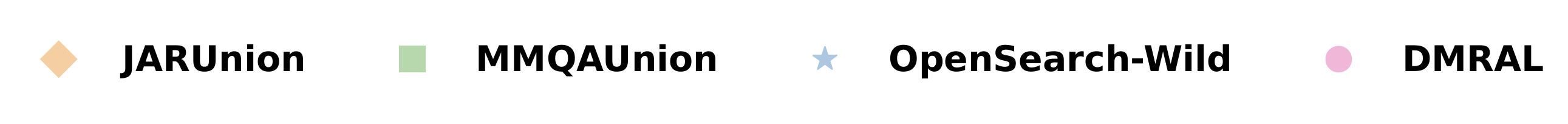}

  \begin{subfigure}[b]{0.5\linewidth}
    \centering
    \includegraphics[width=0.46\linewidth,trim=9mm 10mm 0mm 8mm]{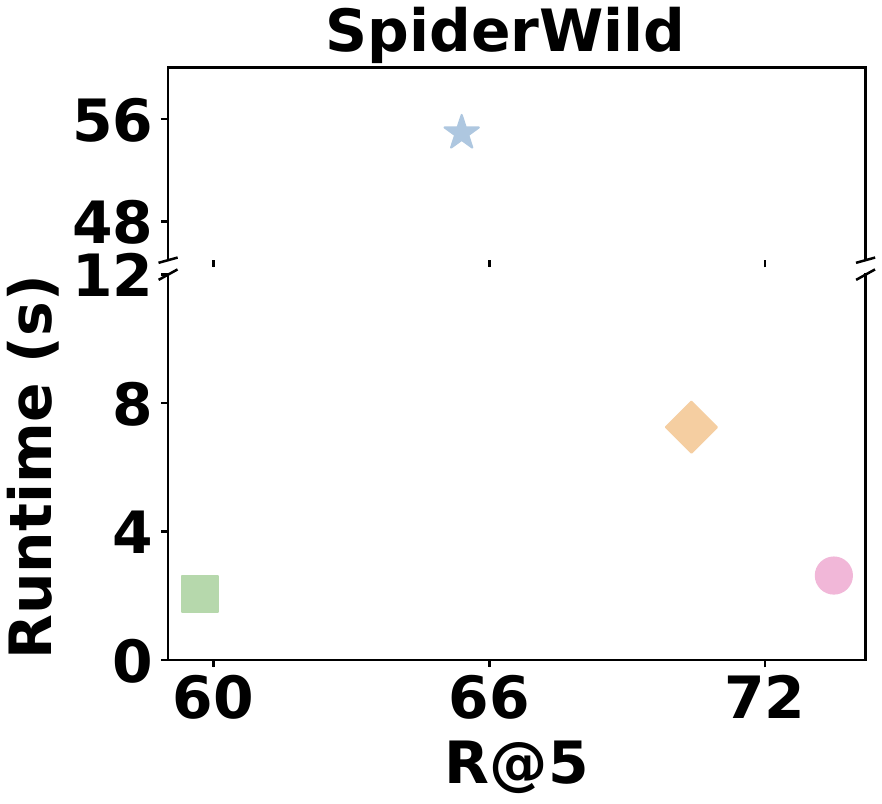}
    \hspace{0.02\linewidth}
    \includegraphics[width=0.44\linewidth,trim=9mm 10mm 0mm 8mm]{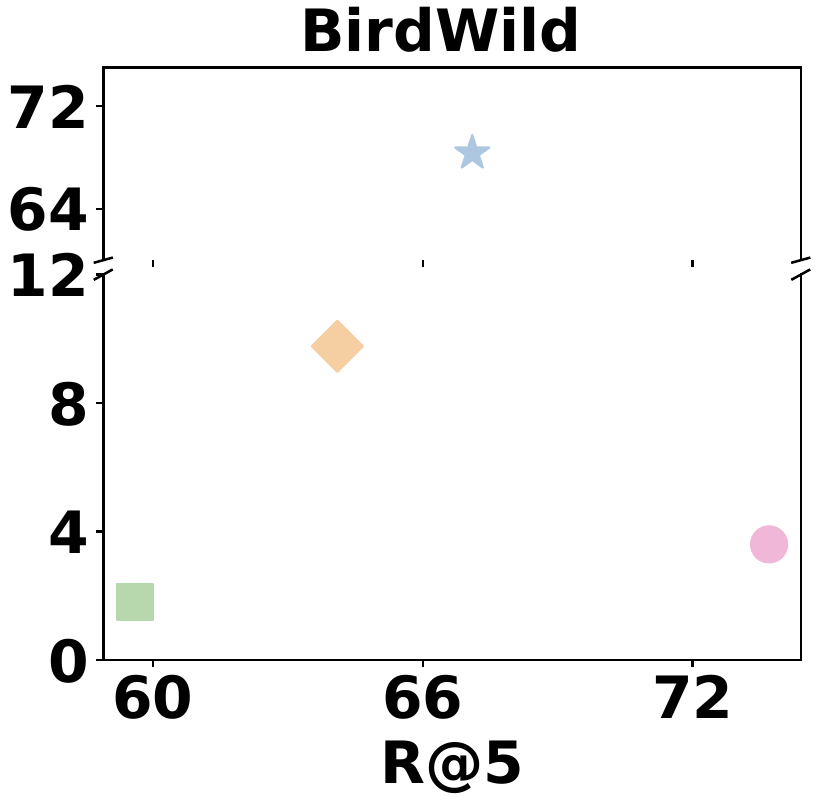}
    \caption{\small Retrieval Efficiency}
    \label{fig:ret_group}
  \end{subfigure}\hfill
  \begin{subfigure}[b]{0.5\linewidth}
    \centering
    \includegraphics[width=0.44\linewidth,trim=9mm 10mm 0mm 8mm]{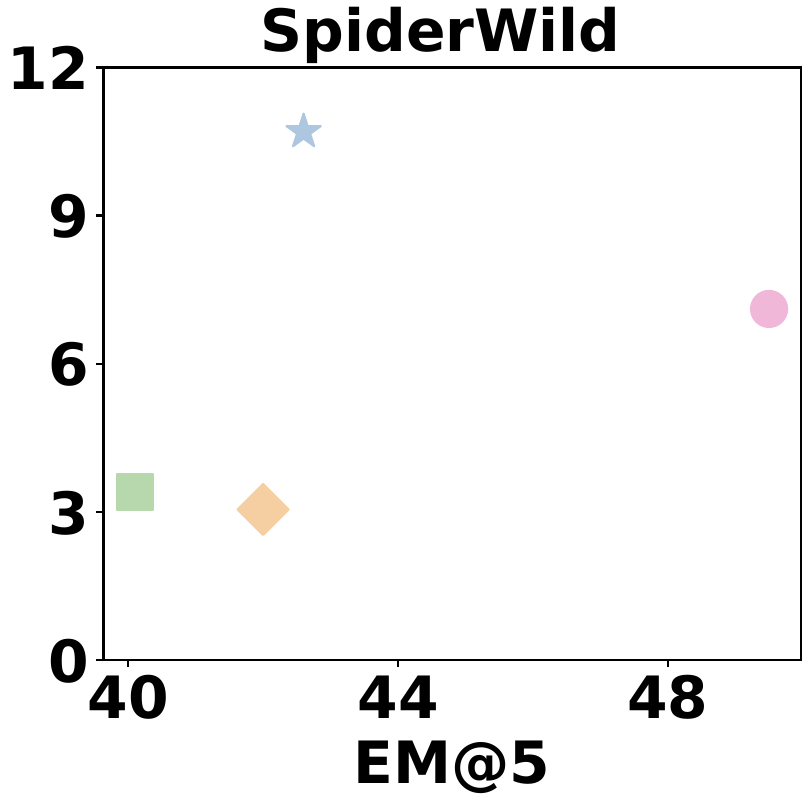}
    \hspace{0.02\linewidth}
    \includegraphics[width=0.44\linewidth,trim=9mm 10mm 0mm 8mm]{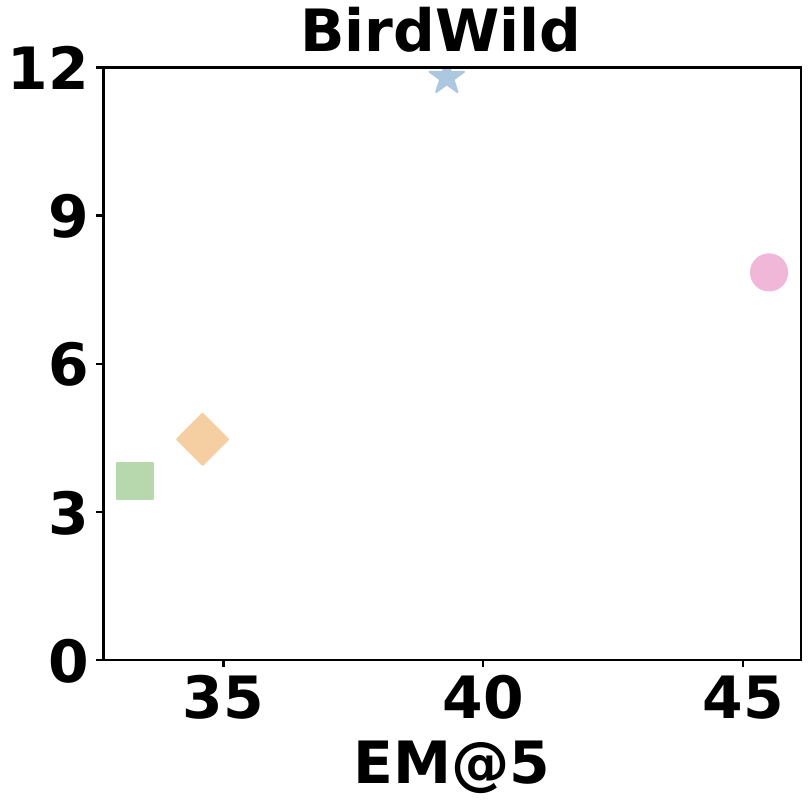}
    \caption{\small Answer Efficiency}
    \label{fig:ans_group}
  \end{subfigure}
\captionsetup{skip=0pt}
  \caption{Efficiency breakdown of table retrieval and answer generation, measured by the average runtime per question.}
  \label{baseline_tradeoff_analysis}
  \vspace{-3mm}
\end{figure}

%% file: ICDE_MTQA/tables/experiments/preprocessing_ablation.tex

\begin{table}[t]
\centering
\small
\setlength{\tabcolsep}{3pt}
\begin{tabular}{llcccc|cccc}
\toprule
\multirow{2}{*}{Top-k} & \multirow{2}{*}{Method} 
& \multicolumn{4}{c}{SpiderWild} 
& \multicolumn{4}{c}{BirdWild} \\
\cmidrule(lr){3-6} \cmidrule(lr){7-10}
& & P & R & F1 & EM  & P & R & F1 & EM \\
\midrule
\multirow{4}{*}{Top-3} & No-fill       & 42.5 & 55.8 & 46.0 & 42.6 & 40.9 & 57.0 & 47.6 & 28.4 \\ 
& Direct LLM    & 50.3 & 67.0 & 57.4 & 44.5 & 48.6 & 66.2 & 56.0 & 40.3 \\
& \ourmethod{}  & 52.3 & 70.3 & 59.4 & 49.0 & 51.5 & 69.5 & 58.6 & 44.3 \\
& GT            & 59.1 & 78.0 & 67.3 & 51.9 & 57.2 & 76.5 & 65.4 & 45.2 \\

\midrule 
\multirow{4}{*}{Top-3} & No-fill       & 25.3 & 59.8 & 35.7 & 43.0 & 26.7 & 58.5 & 36.6 & 29.2 \\ 
& Direct LLM    & 31.4 & 70.0 & 43.4 & 45.0 & 30.9 & 70.2 & 43.0 & 41.4 \\
& \ourmethod{}  & 32.7 & 73.5 & 45.0 & 49.5 & 32.8 & 73.7 & 45.0 & 45.5 \\
& GT            & 37.0 & 81.6 & 50.9 & 52.5 & 36.4 & 81.1 & 50.2 & 46.4 \\
\bottomrule
\end{tabular}
\captionsetup{skip=0pt}
\caption{The impact of metadata quality to the effectiveness of table retrieval and answer generation.}
\label{tab:retrieval_answering_fill}
\vspace{-3mm}
\end{table}

%% file: ICDE_MTQA/tables/experiments/decomposition_metrics.tex
\begin{table}[t]
\centering
\small
\setlength{\tabcolsep}{1pt}
\begin{tabular}{lccc|ccc}
\toprule
\multirow{2}{*}{Method}
& \multicolumn{3}{c}{SpiderWild}
& \multicolumn{3}{c}{BirdWild} \\
\cmidrule(lr){2-4} \cmidrule(lr){5-7}
& IRR~($\uparrow$) & SR~($\downarrow$) & SAR~($\uparrow$) 
& IRR~($\uparrow$) & SR~($\downarrow$) & SAR~($\uparrow$) \\
\midrule
\textit{Direct LLM}           & 93\%          & 0.589 & 57\%           & 94\%          & 0.600 & 59\% \\
\ourmethod{}        & 96\% & 0.521 & 71\% & 96\% & 0.522 & 70\% \\
\bottomrule
\end{tabular}
\captionsetup{skip=0pt}
\caption{Comparison of decomposition quality. $\uparrow$: higher is better, $\downarrow$: lower is better.}
\label{tab:decomposition_quality}
\vspace{-3mm}
\end{table}

%% file: ICDE_MTQA/tables/experiments/decomposition_performance.tex
\begin{table}[t]
\centering
\small
\setlength{\tabcolsep}{3pt}
\begin{tabular}{llcccc|cccc}
\toprule
\multirow{2}{*}{Top-k} &
\multirow{2}{*}{Method} 
& \multicolumn{4}{c}{SpiderWild} 
& \multicolumn{4}{c}{BirdWild} \\
\cmidrule(lr){3-6} \cmidrule(lr){7-10}
& P & R & F1 & EM  & P & R & F1 & EM \\
\midrule
\multirow{2}{*}{Top-3} & \textit{Direct LLM}  & 50.9 & 68.3 & 57.8 & 45.4  & 49.5 & 67.2 & 56.5 & 38.9 \\
  &  \ourmethod{}        & \textbf{52.3} & \textbf{70.3} & \textbf{59.4} & \textbf{49.0}  & \textbf{51.5} & \textbf{69.5} & \textbf{58.6} & \textbf{44.3} \\
\midrule
\multirow{2}{*}{Top-5} & \textit{Direct LLM}  & 31.5 & 70.7 & 43.2 & 49.1  & 32.0 & 72.1 & 43.9 & 39.6 \\
 & \ourmethod{}        & \textbf{32.7} & \textbf{73.5} & \textbf{45.0} & \textbf{49.5}  & \textbf{32.8} & \textbf{73.7} & \textbf{45.0} & \textbf{45.5} \\
\bottomrule
\end{tabular}
\captionsetup{skip=0pt}
\caption{Effectiveness comparison using our decomposer vs. direct LLM-decomposer.}
\label{tab:retrieval_and_answering}
\vspace{-3mm}
\end{table}

%% file: ICDE_MTQA/tables/experiments/retriever_ablation.tex
\begin{table}[t]
\centering
\small
\setlength{\tabcolsep}{2pt}
\begin{tabular}{lcc|cc}
\toprule
\multirow{2}{*}{Method}
& \multicolumn{2}{c}{SpiderWild} 
& \multicolumn{2}{c}{BirdWild} \\
\cmidrule(lr){2-3} \cmidrule(lr){4-5}
& R@3 & R@5 & R@3 & R@5 \\
\midrule
\ourmethod{}        & \textbf{70.3}  & \textbf{73.5}  & \textbf{69.5}  & \textbf{73.7} \\
\textit{NaiveScoring}        & 62.6 {\scriptsize($-$11)} & 66.7 {\scriptsize($-$9)}  & 66.4 {\scriptsize($-$4)}  & 71.3 {\scriptsize($-$3)} \\
\textit{w/o Verification}    & 56.3 {\scriptsize($-$20)} & 60.7 {\scriptsize($-$17)} & 55.9 {\scriptsize($-$20)} & 64.7 {\scriptsize($-$12)} \\
\bottomrule
\end{tabular}
\captionsetup{skip=0pt}
\caption{Ablation study of \methodrt{}\protect\footnotemark. }
\label{tab:coverage_aware_ablation}
\vspace{-5mm}
\end{table}
\footnotetext{Values in parentheses indicate relative performance drops compared to our full method.}

%% file: ICDE_MTQA/tables/experiments/reasoner_ablation.tex
\begin{table}[t]
\centering
\small
\setlength{\tabcolsep}{2pt}
\begin{tabular}{lcc|cc}
\toprule
\multirow{2}{*}{Method} 
& \multicolumn{2}{c}{SpiderWild} 
& \multicolumn{2}{c}{BirdWild} \\
\cmidrule(lr){2-3} \cmidrule(lr){4-5}
& EM@3 & EM@5 & EM@3 & EM@5 \\
\midrule
\ourmethod{}        & \textbf{49.0} & \textbf{49.5}  & \textbf{44.3} & \textbf{45.5} \\
\textit{w/o CoT}    & 44.8 {\scriptsize($-$10)} & 45.2 {\scriptsize($-$10)} & 34.1 {\scriptsize($-$23)} & 37.0 {\scriptsize($-$19)} \\
\textit{w/o Refinement} & 48.3 {\scriptsize($-$3)} & 49.7 {\scriptsize($-$1)} & 37.8 {\scriptsize($-$10)} & 42.0 {\scriptsize($-$8)} \\
\bottomrule
\end{tabular}
\captionsetup{skip=0pt}
\caption{Ablation study of Our Reasoner.}
\label{tab:reasoning_ablation}
\vspace{-5mm}
\end{table}

%% file: ICDE_MTQA/tables/experiments/impact_params.tex
\begin{figure*}[t]
  \centering
  \includegraphics[width=0.24\linewidth]{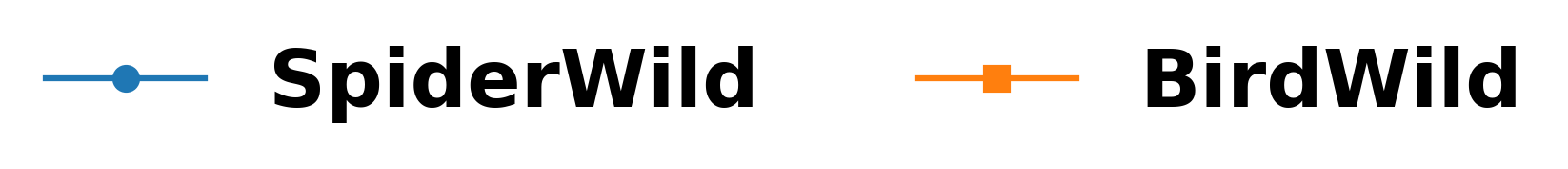}

  \begin{minipage}[t]{0.66\linewidth}
    \centering
    \includegraphics[width=0.24\linewidth,trim=0mm 10mm 0mm 8mm]{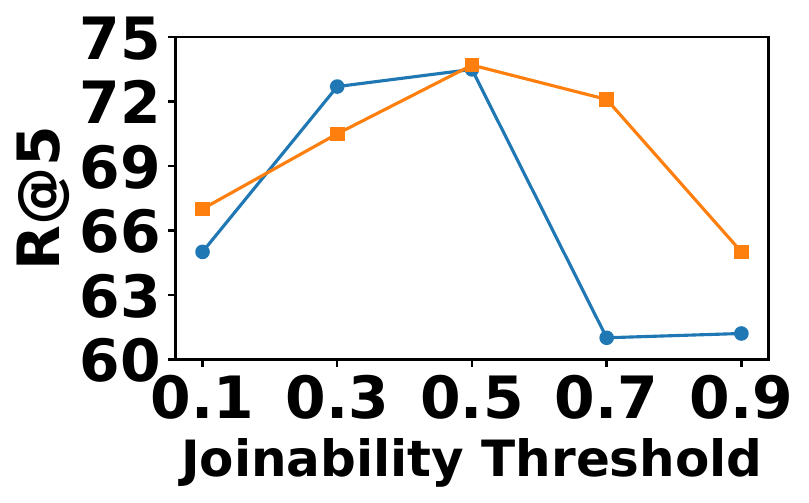}
    \includegraphics[width=0.24\linewidth,trim=0mm 10mm 0mm 8mm]{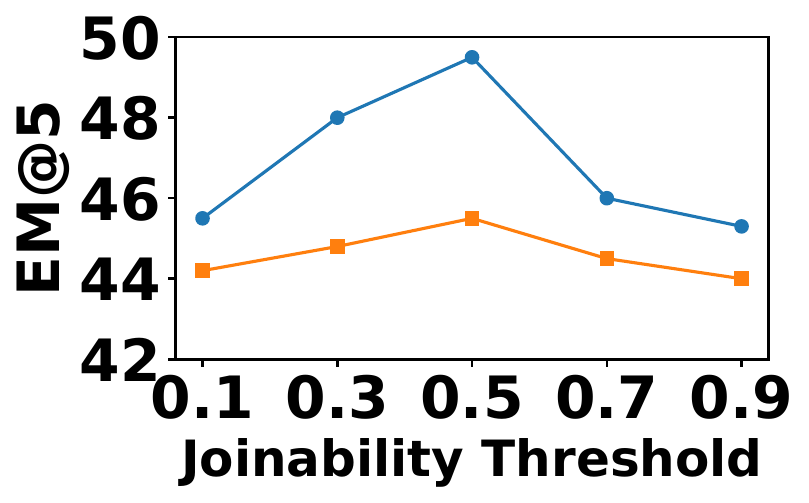}
    \includegraphics[width=0.24\linewidth,trim=0mm 10mm 0mm 8mm]{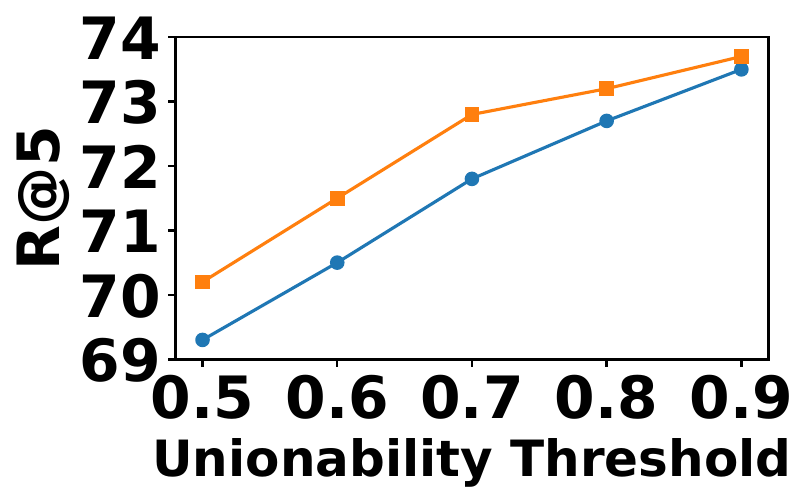}
    \includegraphics[width=0.24\linewidth,trim=0mm 10mm 0mm 8mm]{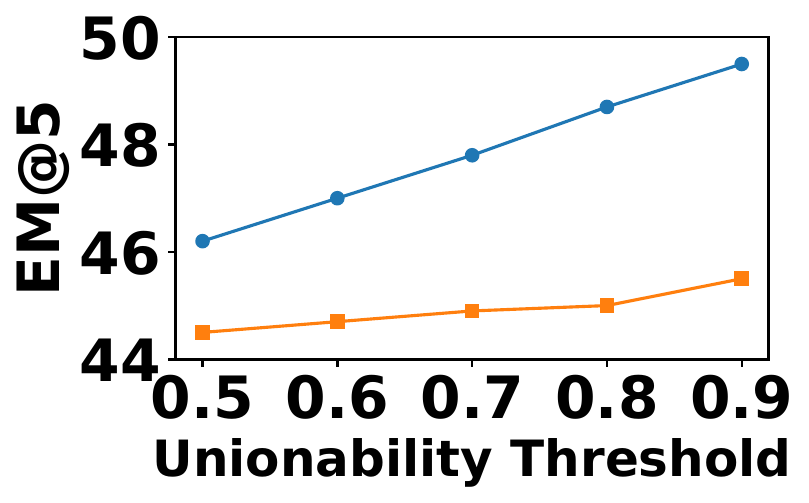}

    \captionof{figure}{Impact of joinability/unionability thresholds on retrieval effectiveness and answer accuracy.}
    \label{fig:impact_param_combined}
  \end{minipage}\hfill
  \begin{minipage}[t]{0.33\linewidth}
    \centering
    \includegraphics[width=0.48\linewidth,trim=0mm 10mm 0mm 8mm]{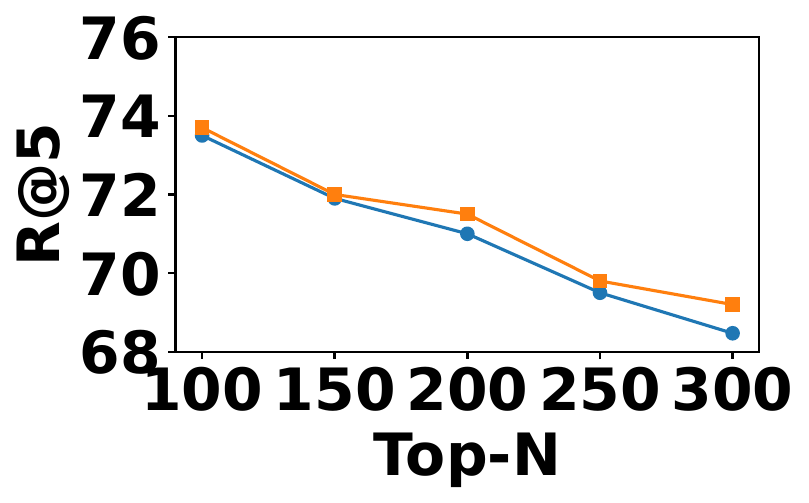}
    \includegraphics[width=0.48\linewidth,trim=0mm 10mm 0mm 8mm]{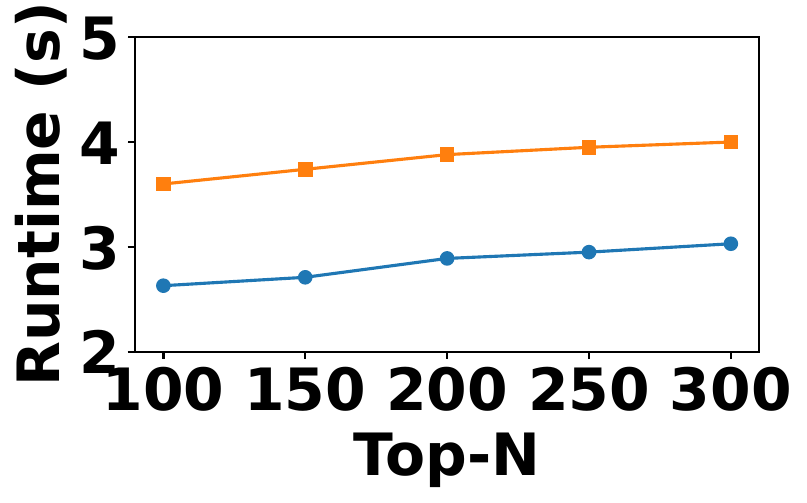}

    \captionof{figure}{Retrieval effectiveness/efficiency vs.\ table scale for \ourmethod{}.}
    \label{fig:impact_scale_test_compact}
  \end{minipage}

  \vspace{-3mm}
\end{figure*}

%% file: ICDE_MTQA/tables/experiments/impact_reltabcnt.tex
    

\begin{figure}[t]
  \centering

  \includegraphics[width=0.9
  \linewidth]{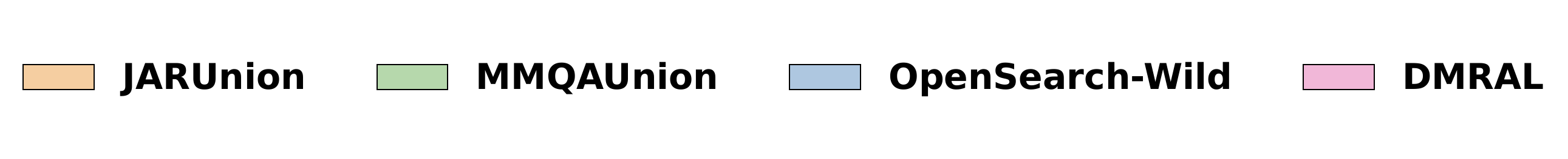}

  \begin{subfigure}[b]{0.48\linewidth}
    \centering
    \includegraphics[width=0.48\linewidth,trim=9mm 10mm 0mm 8mm]{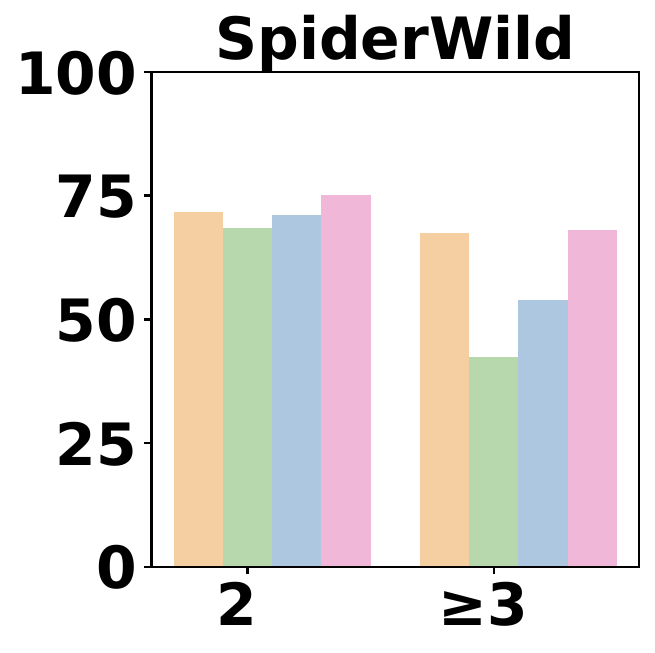}
    \hfill
    \includegraphics[width=0.48\linewidth,trim=9mm 10mm 0mm 8mm]{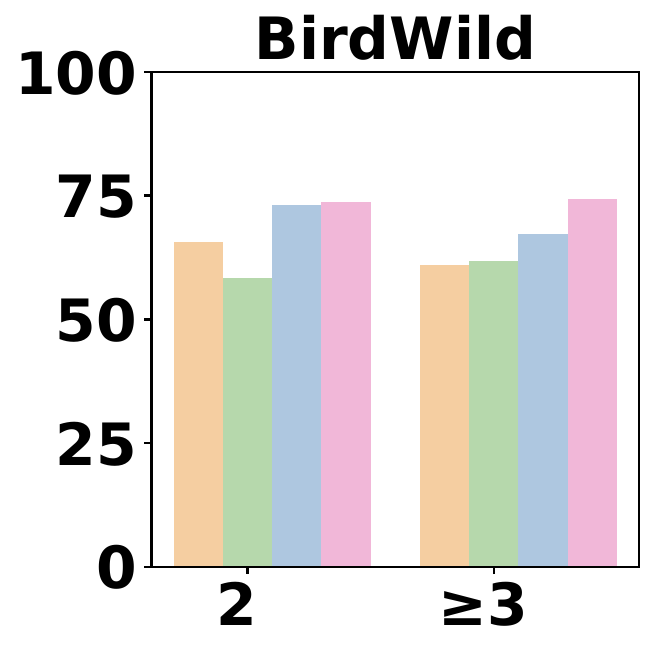}
    \caption{\small Retrieval: R@5}
    \label{reltab_cnt_retrieval}
  \end{subfigure}\hfill
  \begin{subfigure}[b]{0.48\linewidth}
    \centering
    \includegraphics[width=0.48\linewidth,trim=9mm 10mm 0mm 8mm]{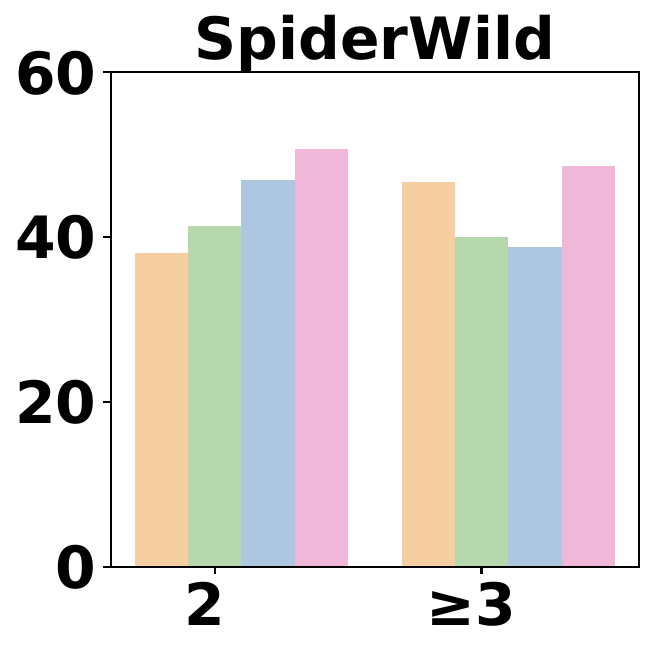}
    \hfill
    \includegraphics[width=0.48\linewidth,trim=9mm 10mm 0mm 8mm]{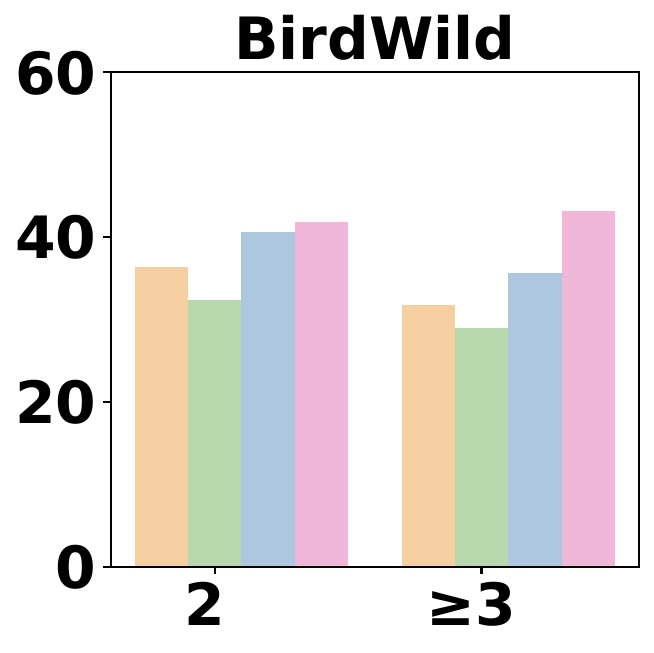}
    \caption{\small Answer: EM@5}
    \label{reltab_cnt_answer}
  \end{subfigure}
    \captionsetup{skip=0pt}
  \caption{Effect of the number of relevant tables (2 vs.\ $\geq$3) on effectiveness.}
  \label{tab:reltabcnt_impact}
  \vspace{-3mm}
\end{figure}

%% file: ICDE_MTQA/tables/experiments/impact_metadata.tex
    

\begin{figure}[t]
  \centering

  \begin{subfigure}[b]{0.48\linewidth}
    \centering
    \includegraphics[width=0.48\linewidth,trim=9mm 10mm 0mm 8mm]{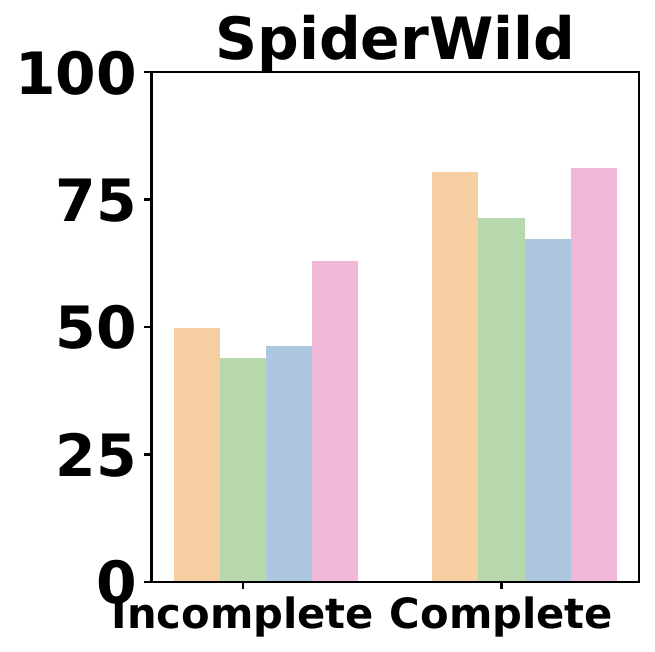}
    \hfill
    \includegraphics[width=0.48\linewidth,trim=9mm 10mm 0mm 8mm]{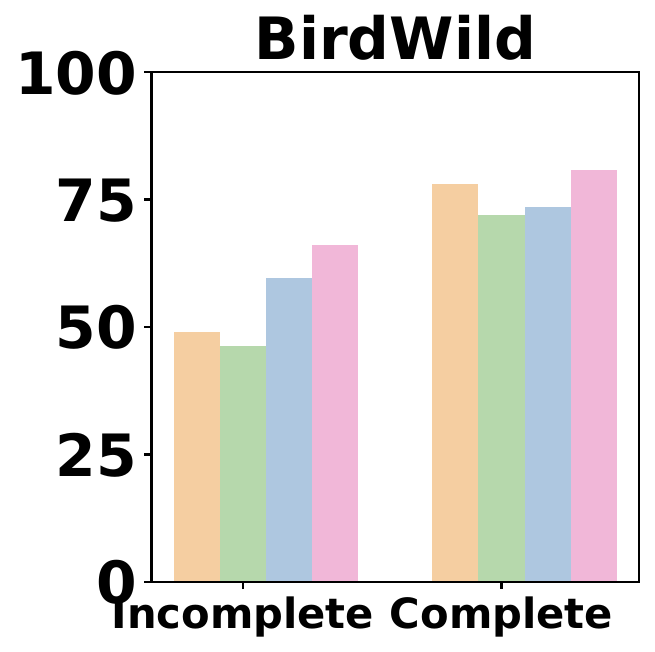}
    \caption{\small Retrieval: R@5}
    \label{metadata_retrieval}
  \end{subfigure}\hfill
  \begin{subfigure}[b]{0.48\linewidth}
    \centering
    \includegraphics[width=0.48\linewidth,trim=9mm 10mm 0mm 8mm]{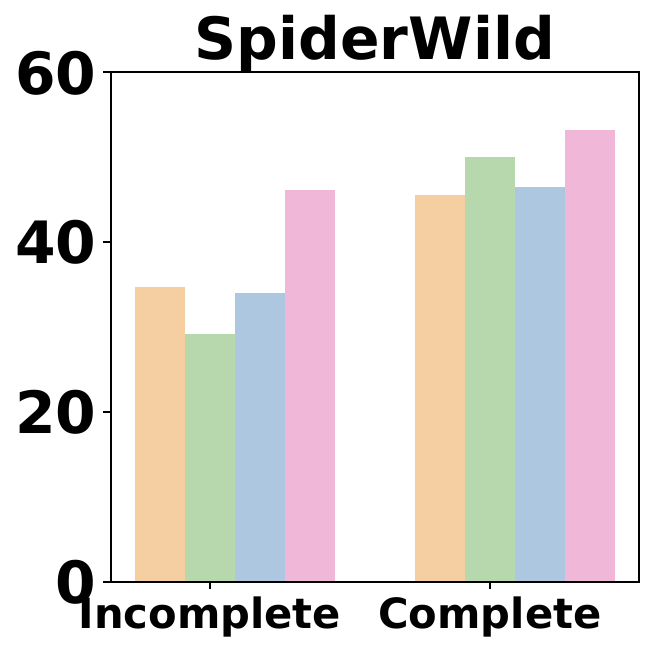}
    \hfill
    \includegraphics[width=0.48\linewidth,trim=9mm 10mm 0mm 8mm]{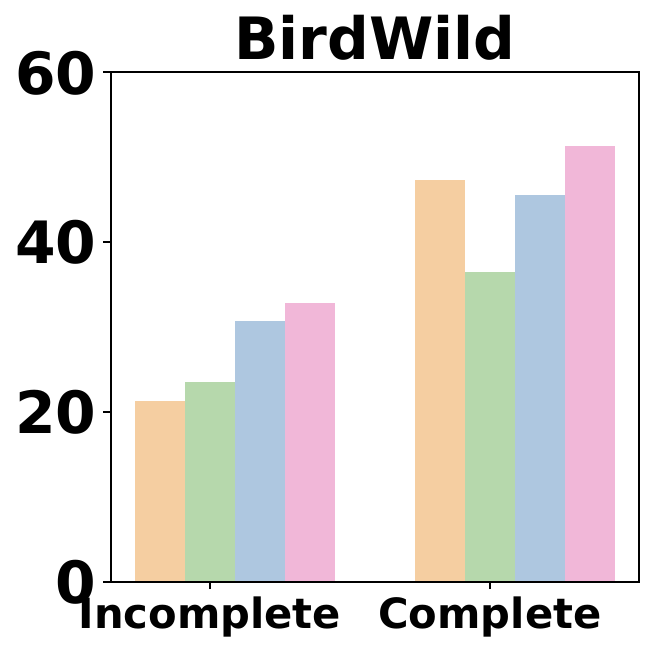}
    \caption{\small Answer: EM@5}
    \label{metadata_answer}
  \end{subfigure}
    \captionsetup{skip=0pt}
  \caption{Evaluation on questions involving tables with incomplete vs.\ complete metadata.}
  \label{tab:metadata_impact}
  \vspace{-5mm}
\end{figure}

%% file: ICDE_MTQA/tables/experiments/impact_union.tex
    

\begin{figure}[t]
  \centering

  \begin{subfigure}[b]{0.48\linewidth}
    \centering
    \includegraphics[width=0.48\linewidth,trim=9mm 10mm 0mm 8mm]{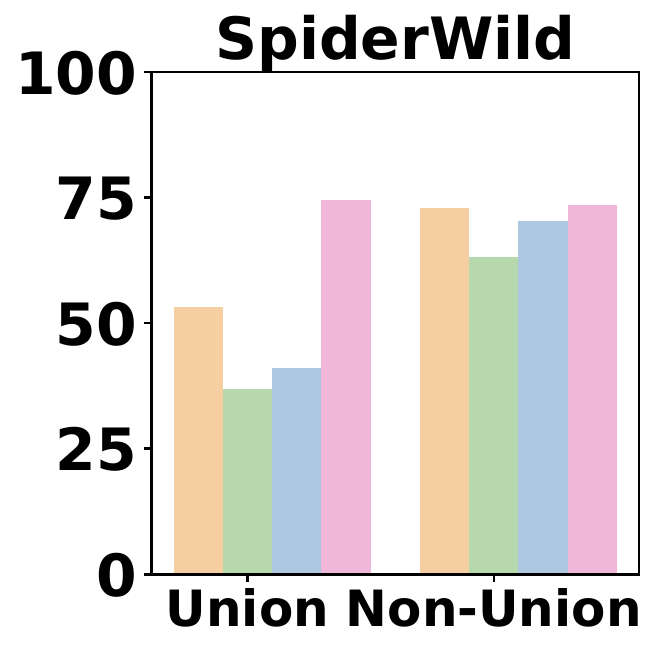}
    \hfill
    \includegraphics[width=0.48\linewidth,trim=9mm 10mm 0mm 8mm]{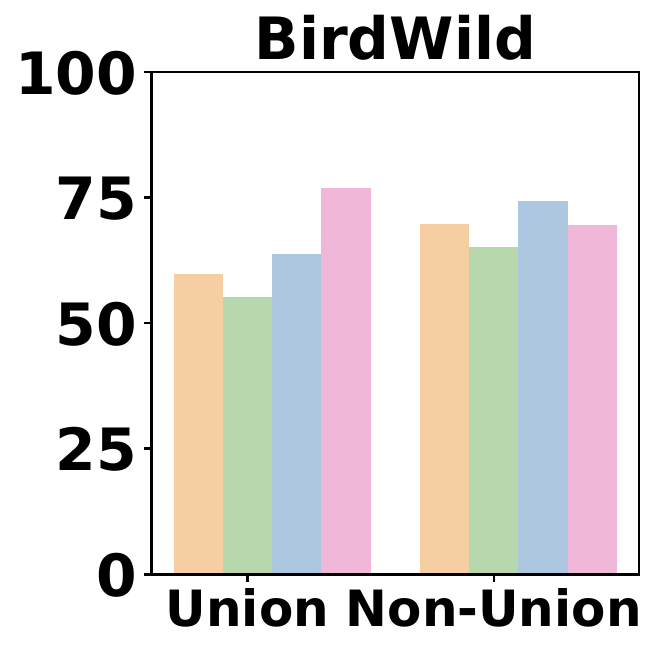}
    \caption{\small Retrieval: R@5}
    \label{union_retrieval}
  \end{subfigure}\hfill
  \begin{subfigure}[b]{0.48\linewidth}
    \centering
    \includegraphics[width=0.48\linewidth,trim=9mm 10mm 0mm 8mm]{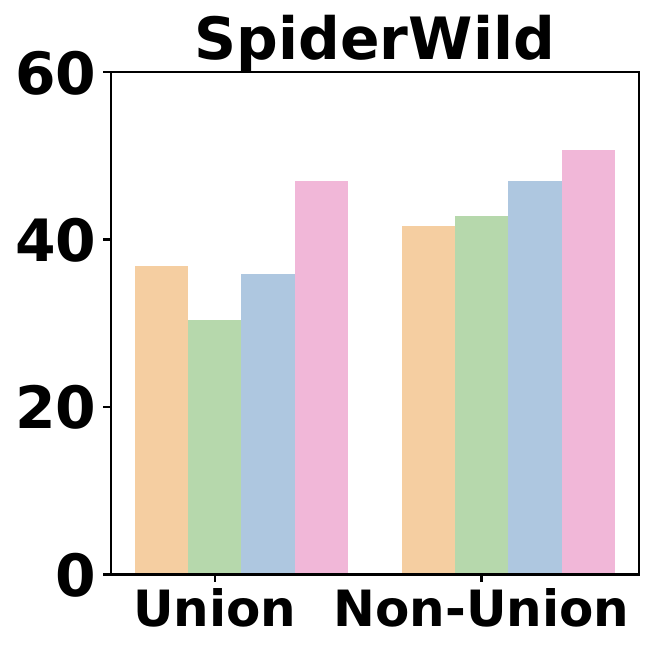}
    \hfill
    \includegraphics[width=0.48\linewidth,trim=9mm 10mm 0mm 8mm]{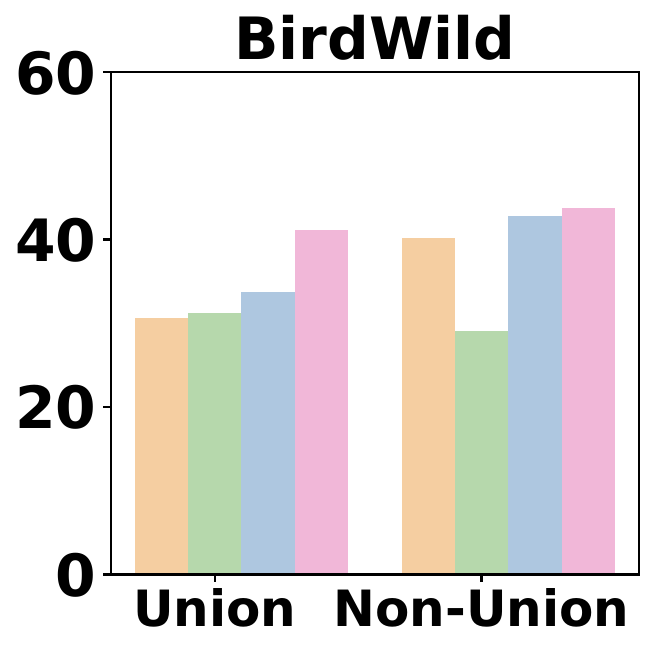}
    \caption{\small Answer: EM@5}
    \label{union_answer}
  \end{subfigure}
    \captionsetup{skip=0pt}
  \caption{Evaluation on questions that require integrating unionable tables (Union) versus those that do not (Non-Union).}
  \label{tab:union_impact}
  \vspace{-5mm}
\end{figure}

%% file: ICDE_MTQA/tables/experiments/text2sql_comparison.tex
\begin{table}[t]
\small
\centering
\setlength{\tabcolsep}{1pt}
\begin{tabular}{lcccccc}
\toprule
\multirow{2}{*}{Method} & \multicolumn{2}{c}{Spider} & \multicolumn{2}{c}{Bird-dev} & \multicolumn{2}{c}{ScienceBenchmark} \\
\cmidrule(lr){2-3}\cmidrule(lr){4-5}\cmidrule(lr){6-7}
 & Table Sel. & Ans. & Table Sel. & Ans. & Table Sel. & Ans. \\
\midrule
DTR & 0.586 & 0.490 & 0.422 & 0.400 & 0.200 & 0.129 \\
TableRAG & 0.667 & 0.597 & 0.600 & 0.267 & 0.294 & 0.118 \\
\midrule
CHESS & 0.769 & 0.801 & 0.733 & 0.733 & 0.324 & 0.306 \\
OpenSearch-SQL & 0.777 & 0.822 & \textbf{0.880} & \textbf{0.867} & 0.482 & 0.400 \\
\midrule
\ourmethod{} & \textbf{0.864} & \textbf{0.826} & 0.867 & 0.778 & \textbf{0.500} & \textbf{0.441} \\
\bottomrule
\end{tabular}
\captionsetup{skip=0pt}
\caption{Comparison with MTQA methods under their original settings, using table selection and answer accuracy.}
\label{tab:text2sql_results}
\vspace{-5mm}
\end{table}

%% file: ICDE_MTQA/5_conclusion.tex
\section{Conclusion}
In this paper, we propose \ourmethod{}, a novel decomposition-driven multi-table retrieval and answering framework designed for numerical MTQA over \settingdef{}. 
\ourmethod{} operates with four modules:
Preprocessing, which constructs a graph to capture complex relationships among tables;
\methodqa{} and \methodrt{}, which jointly improve retrieval by enhancing the quality of decomposition and maximizing the coverage;
\methodrs{} improves the answer quality using guided LLMs to generate an accurate executable program based on decomposed sub-questions.
Experiments using two prepared datasets demonstrate that \ourmethod{} significantly outperforms existing state-of-the-art MTQA methods, achieving an average improvement of $24\%$ in table retrieval and 55\% in answer accuracy.

%% file: ICDE_MTQA/6_appendix.tex
\section{Metadata Inference Module} \label{meta_infer}
As incomplete metadata (i.e., missing column headers) may hinder accurate unionability computation while constructing the graph, we introduce a \textit{metadata inference} module that leverages LLMs to infer and complete the missing headers.
A naive approach here would sample a few rows (e.g., 10) from the table as the whole context and feed them to an LLM to infer missing headers. 
However, this may include irrelevant or loosely correlated columns, which dilute the context and reduce effectiveness \cite{zeng2024multiem}. 
To mitigate this, our module adopts a targeted approach inspired by \cite{sun2023reca}.
The core idea is to utilize intra-table column grouping to establish a concise and semantically coherent context for the LLM, which enhances inference by exclusively focusing the LLM on small, meaningful groups of related columns.
Specifically, we first apply the column semantics discovery algorithm~\cite{khatiwada2023santos} to partition each table into semantically coherent column groups, based on the column values. 
For example, given a table with columns $\text{C} = \{\text{Name, Street Address, City, Product ID, Price, Date of Sale}\}$, the algorithm might partition them into two semantically coherent groups: $\{\text{Name, Street Address, City}\}$ to represent the context of ``Location'', and $\{\text{Product ID, Price, Date of Sale}\}$ under the context of ``Transaction''.
Then, for any column with an incomplete header, we identify its corresponding column group. We then extract only the columns belonging to this group, along with their sampled rows, to construct a subtable. The LLM is subsequently prompted for the header inference using only this highly relevant sub-context (i.e., sampled rows from the partitioned subtable).
The metadata inference prompt is shown in Figure~\ref{prompt_metadata}.

\section{Prompts}  \label{prompt_supp}
In this section, we supplement the main prompts used in this paper.
The question decomposition prompt (\S \ref{sec_ques_decom}) is shown in Figure~\ref{prompt_decomposition}.
The residual sub-question generation prompt (\S \ref{sec:coverage_verification}) is shown in Figure~\ref{prompt_retriever}.
The prompts for multi-step program generation and execution-guided refinement (\S \ref{sec:program_gen}) are shown in Figure~\ref{prompt_cot} and Figure~\ref{prompt_refine}, respectively.
The prompt used to prepare our evaluation datasets by grouping non-key columns into column subsets (\S \ref{sec:data_pre_solution}) is shown in Figure~\ref{promot_col_splitting}.

\section{Details for the adapted competitors} \label{exp_baselines}

\noindent\textbf{JARUnion and MMQAUnion.}
We apply their original table retrieval strategies on our processed metadata-complete tables and augment the retrieved tables by incorporating unionability.
Specifically, for each retrieved table, we expand it by merging all tables in its cluster from our constructed graph to form the final retrieved tables.
We then use their methods on these retrieved tables to obtain answers.

\revise{
\noindent\textbf{OpenSearch-Wild.}
We adapt it to our setting by adding (i) database materialization during its Preprocessing based on our inferred metadata and Table Relationship Graph, and (ii) top-k table selection during its Extraction based on its obtained necessary columns.
Specifically, for materialization, we merge each unionable table cluster into one relational table and construct PK-FK constraints via a lightweight heuristic: we choose the column with the highest uniqueness ratio as a primary key, and for each joinable table pair in our graph we select the column with the highest joinability to the referenced primary key as the foreign key.
For selection, after obtaining the necessary columns for SQL generation, we use an LLM to select the top-k candidate tables from those containing the extracted columns.
}

\section{Error Analysis for Table Retrieval and Answer Generation} \label{exp_error_analysis}

\revise{
We conduct an error analysis on all failure cases in both SpiderWild and BirdWild, including (i) \emph{retrieval errors} where at least one gold relevant table is missed, and (ii) \emph{answer-generation errors} where the generated program yields an incorrect answer. We summarize the dominant error patterns and corresponding improvements below.

\noindent\textbf{Retrieval.}
We observe three dominant retrieval failure patterns: (1) \textit{Missing bridge tables (32\%).} Some questions require an intermediate bridge table to connect two table clusters, but the bridge table is not mentioned and is therefore missed (e.g., joining \texttt{client} and \texttt{card} via \texttt{disp}). 
(2) \textit{Non-informative words (18\%).} Some words mainly provide context but offer little table-relevance signal, which can distract retrieval (e.g.,  ``school'' in the question ``For the school with the highest average score in Reading in the SAT test, what is its FRPM count for students aged 5-17?''). 
(3) \textit{Ambiguous mentions (50\%).} Underspecified entities may be grounded to the wrong tables, causing true evidence tables to be omitted (e.g.,  ``Harlan'' referring to different entity types, leading to missing \texttt{users}).

To mitigate these errors, we plan to strengthen connected table group construction (\S \ref{sec:coverage_verification}) as follows. For \emph{missing bridge tables}, we will expand candidates with 1--2 hop neighbors in the Table Relationship Graph (as bridge table candidates) and rerun group construction and coverage-based reranking; for \emph{non-informative words}, we will apply a post-verification that removes a cluster if top-ranked groups consistently include it and dropping it preserves connectivity with comparable coverage, then update the affected sub-questions; for \emph{ambiguous mentions}, we will incorporate value-level evidence (e.g., matching entities against column values) to verify whether candidate tables truly contain the mentioned entities.

\noindent\textbf{Answer generation.}
We observe three types of program-generation errors:
(1) \textit{Table-selection errors (27\%),} where the reasoner selects incorrect tables or omits required evidence tables;
(2) \textit{Predicate-value grounding errors (27\%),} where it fails to map natural-language value descriptions to the correct table cell values; and
(3) \textit{Operator/structure-level errors (47\%),} where it generates incorrect operators or logical structure (e.g., wrong comparison operator, missing \texttt{DISTINCT}, or inappropriate join types). 

To mitigate these errors, we plan to incorporate three standard Text-to-SQL modules~\cite{xie2025opensearchsqlenhancingtexttosqldynamic} into our CoT-guided multi-step generation and execution (\S \ref{sec:program_gen}): 
(i) \emph{schema linking} during sub-question program generation, we can ground each sub-question to candidate columns first and restricting the table candidates to tables containing these columns, reducing table-selection errors;
(ii) \emph{value retrieval} during predicate construction, we can retrieve a small set of candidate cell values from the selected column (e.g., fuzzy matching over distinct values) and constraining predicate values to these candidates, reducing predicate-value grounding errors; and
(iii) \emph{self-consistency voting} before execution, we can generate multiple candidate programs, execute them, and select the most consistent execution result, reducing operator/structure errors.
}

\section{Case Analysis} \label{exp_supp}
We conduct a qualitative comparison between the subquestions produced by the \textit{Direct LLM} approach and the table-aligned question decomposer in \ourmethod{}.  
Table~\ref{tab:decomposition_issues} presents representative examples highlighting the major decomposition issues commonly observed with the \textit{Direct LLM} approach, alongside the improvements achieved by \ourmethod{}.
We categorize these issues into three types:

\noindent \textbf{Missing Key Information:}  
    Critical elements required to fully specify the question are omitted.  
    For example, in the first case, the \textit{Direct LLM} decomposition misses the mention of \textit{"account opened"}, which is essential to correctly identify the relevant table \texttt{account}.  
    In contrast, \ourmethod{} explicitly retains this information in the second sub-question (i.e., better completeness).

\noindent  \textbf{Redundant Decomposition:}  
    The decomposition produces overlapping or repetitive sub-questions.  
    In the second case, both sub-questions generated by \textit{Direct LLM} redundantly reference \textit{"superheroes"}, resulting in duplicate query intent for table \textit{superhero}. 
    In contrast,  \ourmethod{} avoids such redundancy (i.e., better non-redundancy).

\noindent \textbf{Entangled Sub-question:} 
    The decomposition produces a sub-question that entangles multiple distinct information needs into a single sub-question, increasing the answering complexity.
    In the third case, \textit{Direct LLM} merges \textit{"segment SME"} and \textit{"year 2013"} into a single sub-question, even though both elements are less likely to co-occur in the same context.
    In contrast, \ourmethod{} separates them into two sub-questions (i.e., better table-specificity).

\input{ICDE_MTQA/tables/prompts/metadata}
\input{ICDE_MTQA/tables/prompts/decomposition}
\input{ICDE_MTQA/tables/prompts/col_splitting}
\input{ICDE_MTQA/tables/prompts/retriever}
\input{ICDE_MTQA/tables/prompts/cot_prompt}
\input{ICDE_MTQA/tables/prompts/refine_prompt}

\input{ICDE_MTQA/tables/experiments/case_analysis_decomposition}

%% file: ICDE_MTQA/tables/prompts/metadata.tex
\begin{figure*}[t]
\centering
\small
\begin{tcolorbox}[width=\linewidth]
You are an expert in metadata inference for table-based data understanding.

\vspace{0.5em}
\noindent\textbf{Task:}  
You are provided with a table containing partially missing metadata, such as incomplete or masked table titles and column headers.  
Your goal is to infer and recover the missing metadata based on the provided table structure and inter-table context (i.e., sampled rows).

\vspace{0.5em}
\noindent\textbf{Important Instructions:}\\
\noindent-- You must return exactly the same number of column headers as provided in the "Column Headers".\\
\noindent-- Do not add, remove, or reorder any columns—only replace missing ones with your best inference.\\
\noindent-- Maintain the original header order.

\vspace{0.5em}
\noindent\textbf{Response Format:}
\begin{verbatim}
{
  "updated_title": "...",
  "updated_headers": ["...", "...", ...]
}
\end{verbatim}

\vspace{0.5em}
\noindent\textbf{Now perform the metadata inference on the provided table:}

\vspace{0.5em}
\noindent-- [Table Title] \texttt{\{table\_title\}}\\
\noindent-- [Column Headers] \texttt{\{headers\}}\\
\noindent-- [Sample Rows (10 randomly sampled rows)] \texttt{\{sampled\_rows\}}

\end{tcolorbox}
\caption{Metadata Inference Prompt.}
\label{prompt_metadata}
\end{figure*}

%% file: ICDE_MTQA/tables/prompts/decomposition.tex
\begin{figure*}[t]
\centering
\small
\begin{tcolorbox}[width=\linewidth]

You are an expert in multi-hop question decomposition for table-based question answering.

\noindent\textbf{Task:}  
Decompose a complex question into a sequence of simpler, minimal sub-questions.  
Each sub-question must satisfy the following guidelines:

\noindent-- Use the key phrases grouped together in each entry from a provided list called "information needs".\\
\noindent-- Preserve the full meaning and intent of the original question without omitting important details.\\
\noindent-- Ensure the sub-questions are minimal, non-redundant, and natural.

\noindent\textbf{Response Format:}
\begin{verbatim}
{"Sub-questions": [...]}
\end{verbatim}
\noindent\textbf{Example:}
\noindent [Question]  
Among the schools with the average Math score over 560 in the SAT test, how many schools are directly charter-funded?

\noindent [Information Needs]  
\texttt{[['schools', 'SAT test', 'charter funded'], ['Math score over 560']]}

\noindent Output:
\begin{verbatim}
{"Sub-questions": [ "Which schools have an 
average Math score over 560?",
"How many of the schools from #1 are 
directly charter-funded in the SAT test?"]}
\end{verbatim}

\noindent\textbf{Now decompose the following question:}

\noindent-- [Question] \texttt{\{question\}}\\
\noindent-- [Information Needs] \texttt{\{table-aligned groups\}}

\end{tcolorbox}
\captionsetup{skip=0pt}
\caption{Question Decomposition Prompt.}
\vspace{-3mm}
\label{prompt_decomposition}
\end{figure*}

%% file: ICDE_MTQA/tables/prompts/col_splitting.tex
\begin{figure*}[t]
\centering
\small
\begin{tcolorbox}[width=\linewidth]
You are an expert in organizing tables by grouping related columns into smaller, meaningful subtables.

\vspace{0.5em}
\noindent\textbf{Task:}  
You are provided with a cluttered table and asked to reorganize its columns into subtables.

\vspace{0.5em}
\noindent\textbf{Reorganization Guidelines:}\\
\noindent-- Group columns that naturally belong together.\\
\noindent-- Do not rename, remove, or reorder the columns within each group.

\vspace{0.5em}
\noindent\textbf{Response Format:}
\begin{verbatim}
{
  "Tables": [
    {
      "Table Title": "...",
      "Column Headers": ["...", "...", ...]
    },
    ...
  ]
}
\end{verbatim}

\vspace{0.5em}
\noindent\textbf{Now perform the table reorganization based on the provided table context.}

\vspace{0.5em}
\noindent-- Table Title: \texttt{\{table\_title\}}\\
\noindent-- Column Headers: \texttt{\{column\_headers\}}

\end{tcolorbox}
\caption{Semantic Column Grouping Prompt.}
\label{promot_col_splitting}
\end{figure*}

%% file: ICDE_MTQA/tables/prompts/retriever.tex
\begin{figure*}[t]
\centering
\small
\begin{tcolorbox}[width=\linewidth]
You are an expert in identifying semantic gaps in query decomposition.

\vspace{0.5em}
\noindent\textbf{Task:}  
Identify any missing tables needed to fully answer the question as a residual sub-question, given a user question and a set of currently available tables.  
If all required tables are already provided, return \texttt{None}.

\vspace{0.5em}
\noindent\textbf{Response Format:}
\begin{verbatim}
{"Residual Sub-question": ...}
\end{verbatim}

\vspace{0.5em}
\noindent\textbf{Examples:}

\vspace{0.5em}
\noindent [Question]  
How many female clients opened their accounts in Jesenik branch?

\noindent [Provided Tables]  
\texttt{financial.client(client\_id, gender, birth date, district\_id)}\\
\texttt{financial.disp(disposition\_id, client\_id, account\_id, type)}

\noindent Output:
\begin{verbatim}
{"Residual Sub-question": "What is the 
district name of 
the Jesenik branch?"}
\end{verbatim}

\vspace{0.5em}
\noindent [Question]  
Among the atoms that contain element carbon, which one does not contain compound carcinogenic?

\noindent [Provided Tables]  
\texttt{toxicology.atom(atom\_id, molecule\_id, element)}\\
\texttt{toxicology.molecule(molecule\_id, label)}

\noindent Output:
\begin{verbatim}
{"Residual Sub-question": None}
\end{verbatim}

\vspace{0.5em}
\noindent\textbf{Now perform the task on the following input:}

\vspace{0.5em}
\noindent-- [Question] \texttt{\{question\}}\\
\noindent-- [Provided Tables] \texttt{\{tables\}}

\end{tcolorbox}
\caption{Residual Sub-question Generation Prompt.}
\label{prompt_retriever}
\end{figure*}

%% file: ICDE_MTQA/tables/prompts/cot_prompt.tex
\begin{figure*}[t]
\centering
\small
\begin{tcolorbox}[width=\linewidth]
You are an expert in SQL program synthesis for multi-table question answering over structured tabular data.

\vspace{0.5em}
\noindent\textbf{Task:}  
Your objective is to reason step by step to generate the correct SQL program.  
You are provided with:

\noindent-- A natural language question.\\
\noindent-- A set of retrieved tables with metadata.\\
\noindent-- A set of decomposed sub-questions (which may be inaccurate or incomplete).\\
\noindent-- Optional external knowledge (e.g., mappings between question phrases and columns).

\vspace{0.5em}
\noindent\textbf{Reasoning Process:}\\
\noindent-- \textbf{Step 1:} Carefully read the question and table metadata to understand the full requirements.\\
\noindent-- \textbf{Step 2:} Evaluate the decomposed sub-questions. Revise or rewrite them if needed.\\
\noindent-- \textbf{Step 3:} For each revised sub-question, reason step by step to write the corresponding SQL query.\\
\noindent-- \textbf{Step 4:} After processing all sub-questions, check if the final sub-question’s SQL fully answers the original question.

\vspace{0.5em}
\noindent\textbf{Response Format:}
\begin{verbatim}
{
  "reasoning": "...",
  "Final SQL": "..."
}
\end{verbatim}

\vspace{0.5em}
\noindent\textbf{Example:}

\vspace{0.5em}
\noindent [Question]  
List the names of female students who have enrolled in more than 3 courses.

\noindent [Retrieved Tables]  
\texttt{enrollment(enrollment\_id, student\_id, course\_id)}  
\texttt{student(student\_id, name, age, gender)}

\noindent [External Knowledge]  
"female" refers to gender = 'F'

\noindent [Decomposed Sub-questions]  
\texttt{['Which students have enrolled
in more than 3 courses?', 
'What are the names of female students from \#1?']}

\noindent Output:
\begin{verbatim}
{
  "reasoning": "Let's think step by step.
  Step 1: Revised sub-question 1: 
  Which students have enrolled in more than 3 courses?
  SQL: SELECT student_id FROM enrollment GROUP BY student_id HAVING COUNT(course_id) > 3.

  Step 2: Revised sub-question 2: What are the names of female students from #1?
  SQL: SELECT DISTINCT name FROM student WHERE student_id IN (...) AND gender = 'F'."

  "Final SQL": "SELECT DISTINCT name FROM student WHERE student_id IN (SELECT student_id 
  FROM enrollment GROUP BY student_id HAVING COUNT(course_id) > 3) AND gender = 'F'"
}
\end{verbatim}

\vspace{0.5em}
\noindent\textbf{Now perform the reasoning and SQL program generation on the provided input:}

\vspace{0.5em}
\noindent-- [Question] \texttt{\{question\}}\\
\noindent-- [Retrieved Tables] \texttt{\{tables\}}\\
\noindent-- [External Knowledge] \texttt{\{external\_knowledge\}}\\
\noindent-- [Decomposed Sub-questions] \texttt{\{decomposed\ sub-questions\}}

\end{tcolorbox}
\caption{Chain-of-Thought Guided Program Generation Prompt.}
\label{prompt_cot}
\end{figure*}

%% file: ICDE_MTQA/tables/prompts/refine_prompt.tex
\begin{figure*}[t]
\centering
\small
\begin{tcolorbox}[width=\linewidth]
You are an expert SQL correction assistant for multi-table question answering over structured tabular data.

\vspace{0.5em}
\noindent\textbf{Task:}  
Given a natural language question, a set of related tables with metadata, a faulty SQL query that failed with a SQL execution error, and optional external knowledge (e.g., mappings between question phrases and columns),  
your task is to correct the SQL query by fixing the identified error.

\vspace{0.5em}
\noindent\textbf{Response Format:}
\begin{verbatim}
{
  "SQL": "..."
}
\end{verbatim}

\vspace{0.5em}
\noindent\textbf{Now perform the SQL correction based on the provided context:}

\vspace{0.5em}
\noindent-- [Question] \texttt{\{question\}}\\
\noindent-- [Tables] \texttt{\{tables\}}\\
\noindent-- [Error SQL] \texttt{\{sql\}}\\
\noindent-- [SQL Error] \texttt{\{sql\_error\}}\\
\noindent-- [External Knowledge] \texttt{\{external\_knowledge\}}

\end{tcolorbox}
\caption{Program Refinement Prompt.}
\label{prompt_refine}
\end{figure*}

%% file: ICDE_MTQA/tables/experiments/case_analysis_decomposition.tex
\begin{table*}[b]
\centering
\small
\begin{tabular}{|>{\centering\arraybackslash}p{2cm}|
                >{\centering\arraybackslash}p{3.2cm}|
                >{\centering\arraybackslash}p{4.5cm}|
                >{\centering\arraybackslash}p{4.5cm}|}
\hline
Issue & Question & \textit{Direct LLM} & \ourmethod{} \\
\hline

\textit{Missing Key Information}& \textit{Among the \textcolor{red}{account opened}, how many female customers born before 1950 and stayed in Sokolov?}& 
SubQ1: Which female customers were born before 1950?  
SubQ2: How many of \#1 stayed in Sokolov?& 
SubQ1:  Which female customers were born before 1950 and stayed in Sokolov?  
SubQ2: How many of \#1 opened an account?\\
\hline

\textit{Redundant Decomposition}& \textit{How many superheroes have brown eyes?} & 
SubQ1: Which \textcolor{red}{superheroes} have brown eyes?  
SubQ2: How many of the \textcolor{red}{superheroes} from \#1?& 
SubQ1: Which are brown eyes?  
SubQ2: How many superheroes have \#1?\\
\hline

\textit{Entangled Sub-question}& \textit{What was the average monthly consumption of customers in segment SME for the year 2013?}& 
SubQ1: Which customers are in \textcolor{red}{segment SME} for the \textcolor{red}{year 2013}?  
SubQ2: What was the average monthly consumption for  \#1?& 
SubQ1: Which customers are in segment SME?
SubQ2: What was the average monthly consumption for  \#1 in year 2013?\\
\hline
\end{tabular}
\caption{Examples of decomposition issues observed in \textit{Direct LLM} vs. \ourmethod{}.}
\label{tab:decomposition_issues}
\end{table*}